\documentclass[12pt,preprint]{aastex}

\slugcomment{}

\shorttitle{Dispersing Envelope in L1551 NE}
\shortauthors{Takakuwa et al.}

\begin{document}


\title{Dispersing Envelope around the Keplerian Circumbinary Disk\\
in L1551 NE and its Implications for the Binary Growth}


\author{Shigehisa Takakuwa\altaffilmark{1}, Kazuhiro Kiyokane\altaffilmark{2},
Kazuya Saigo\altaffilmark{3}, \& Masao Saito\altaffilmark{4}}
\altaffiltext{1}{Academia Sinica Institute of Astronomy and Astrophysics,
P.O. Box 23-141, Taipei 10617, Taiwan; takakuwa@asiaa.sinica.edu.tw}
\altaffiltext{2}{Department of Astronomy, Graduate School of Science, The University of Tokyo, Hongo 7-3-1, Bunkyo-ku, Tokyo 113-0033, Japan}
\altaffiltext{3}{Department of Physical Science, Graduate School of Science, Osaka Prefecture University,
1-1 Gakuen-cho, Naka-ku, Sakai, Osaka 599-8531, Japan}
\altaffiltext{4}{Nobeyama Radio Observatory, National Astronomical Observatory of Japan,
Minamimaki, Minamisaku, Nagano 384-1805, Japan}


\begin{abstract}
We performed mapping observations of the Class I protostellar binary system L1551 NE in the C$^{18}$O ($J$=3--2), $^{13}$CO ($J$=3--2), CS ($J$=7--6), and SO ($J_N$=7$_8$--6$_7$) lines with Atacama Submillimeter Telescope Experiment (ASTE). The ASTE C$^{18}$O data are combined with our previous SMA C$^{18}$O data, which show a $r \sim$300-AU scale Keplerian disk around the protostellar binary system. The C$^{18}$O maps show a $\sim$20000-AU scale protostellar envelope surrounding the central Keplerian circumbinary disk. The envelope exhibits a northeast (blue) - southwest (red) velocity gradient along the minor axis, which can be interpreted as a dispersing gas motion with an outward velocity of 0.3 km s$^{-1}$, while no rotational motion in the envelope is seen. In addition to the envelope, two $\lesssim$4000 AU scale, high-velocity ($\gtrsim$1.3 km s$^{-1}$) redshifted $^{13}$CO and CS emission components are found to $\sim$40$\arcsec$ southwest and $\sim$20$\arcsec$ west of the protostellar binary. These redshifted components are most likely outflow components driven from the neighboring protostellar source L1551 IRS 5, and are colliding with the envelope in L1551 NE. The net momentum, kinetic and internal energies of the L1551 IRS 5 outflow components are comparable to those of the L1551 NE envelope, and the interactions between the outflows and the envelope are likely to cause the dissipation of the envelope and thus suppression of the further growth of the mass and mass ratio of the central protostellar binary in L1551 NE.
\end{abstract}





\section{Introduction}

The physical mechanisms
to set the mass ratios of binary stars is of fundamental astrophysical importance,
since more than half of main-sequence, pre-main sequence, and even protostars having masses
comparable to the Sun are members of binary systems \cite{rag10,che13,rei14}.
Binary stars with the primary masses comparable to the solar mass exhibit a wide,
essentially flat distribution of the mass ratios ($\equiv q$) from $q \sim$0.1 to 1 \cite{rag10,goo13}.
The physical processes to reproduce such a flat distribution of the binary mass ratios are,
however, still controversial.

Protostellar binary systems, precursors of main-sequence binary systems,
are often surrounded by disks of molecular gas
and dusts, ``circumbinary disks'' (hereafter CBDs) \cite{tak12,tob13,cho14,dut14,tak14,tan14}.
The radii of these CBDs range from $\lesssim$100 to $\sim$500 AU, and these CBDs often show
inner emission depressions and ringlike structures.
Mass accretion from these CBDs onto the central protostellar binaries has been considered as one of the
key physical processes to set the binary mass ratios.
Recent high-resolution observations of the CBDs have indeed detected gas flows from CBDs to the inner
protostellar binaries, while the spatial resolutions are not high enough to disentangle the accretions
onto the primaries and secondaries \cite{dut14,tak14,tan14}.
Theoretically,
Bate \& Bonnell (1997) and Bate (2000) have shown that from their 
Smoothed Particle Hydrodynamic (SPH) simulations the majority of
circumbinary materials accrete onto the secondary.
High-resolution grid-based simulations by Ochi et al. (2005) and Hanawa et al. (2010) have shown
the opposite results, that is, the primary accretes more than the secondary.
Ochi et al. (2005) and Hanawa et al. (2010) argued that high-resolution simulations
to sufficiently resolve the gas motions near the L2 and L3 Lagrangian points
are essential to properly trace the mass accretion, since accreting gas flows into the binary
through these Lagrangian points.
On the other hand, latest SPH simulations have demonstrated that mass accretion in CBDs
is a function of gas temperatures, and
in the case of the high temperature the primary accretes more than the secondary,
and vice versa \cite{you15}. Young et al. (2015) argued that the difference between the above results
can simply be explained by the difference of the adopted gas temperatures in the models.
It has been still controversial to construct a consistent, unified model of mass accretion from
CBDs onto protostellar binaries.

CBDs in protostellar binary systems are considered
to be embedded in larger-scale protostellar envelopes,
which can replenish the central CBDs with fresh materials \cite{mom98,cho14}.
Since the observed masses of CBDs around protostellar binaries ($\lesssim$0.1 $M_{\odot}$)
are much smaller than the masses of the central protostellar binaries
\cite{tak12,tob13,cho14},
further supply of materials from the protostellar envelopes to the CBDs is essential
to significantly change the mass and mass ratios of the protostellar binaries.
If a protostellar envelope keeps supplying materials to the central CBD,
further growth of
the mass and the mass ratio of the protostellar binary will be expected.
On the other hand, if the most of the envelope materials are being dissipated,
the present mass and mass ratio of the protostellar binary will be close to the final values.
Such dissipations of protostellar envelopes through the interactions with the associated outflows
or stellar winds have been investigated observationally \cite{kit96,mom96,fue02,tak03,arc06,tak06}.
Thus, in addition to physics of accretion from CBDs onto the
protostellar binaries, mass replenishment from the surrounding protostellar envelopes,
and the connection between protostellar envelopes and CBDs,
must be taken into account.

In the present paper, we focus on a protostellar envelope harboring
the protostellar binary system L1551 NE. L1551 NE is a young Class I protostellar
binary ($T_{bol}$=91 K, $L_{bol}$=4.2 $L_{\odot}$; Froebrich 2005)
located to $\sim$2$\farcm$5 northeast of another, brightest protostellar binary L1551 IRS 5
in the L1551 region \cite{sai01,rei02,tak04,hay09,cho14}.
The protostellar binary consists of the northwestern source named ``Source A'',
and the southeastern source ``Source B'', with the projected separation of $\sim$70 AU
at a position angle of 300$\degr$ \cite{rei02,tak14}.
Our previous SMA observations of L1551 NE
have identified a $r \sim$300 AU scale CBD in Keplerian rotation
with a central stellar mass of 0.8 $M_{\odot}$ \cite{tak12}, plus a possible
outer infalling component \cite{tak13}.
Our subsequent ALMA Cycle 0 observation of L1551 NE \cite{tak14},
at a spatial resolution that is $\sim$1.6 times higher (in beam
area) and a sensitivity that is $\sim$6 times better (in brightness
temperature) than those attained in our previous SMA observations,
unveiled substructures of the CBD in the 0.9-mm dust-continuum emission.
The revealed substructures are
consistent with our 3-D adaptive mesh refinement (AMR) hydrodynamic simulation \cite{mat07} 
for the presence of two spiral arms driven by
gravitational torques from the central binary system. The ALMA data of L1551 NE
in the C$^{18}$O (3--2) line also exhibits the deviations from
the Keplerian motion in the CBD, consistent with our AMR simulation that
gravitational torques impart angular momentum along the spiral arms (driving material outwards) and
extract angular momentum between the spiral arms (driving infall).
Our theoretical model which reproduces the observed features of the CBD around L1551 NE
predicts that secondary accretes more than the primary.
Since our series of the SMA and ALMA observations of L1551 NE have revealed the CBD + protostellar binary system,
the next question is the connection from the protostellar envelope to the CBD, and
the mass replenishment from the envelope to the CBD.

We conducted single-dish mapping observations of
the protostellar envelope surrounding the protostellar binary system of L1551 NE
in the C$^{18}$O ($J$=3--2), $^{13}$CO ($J$=3--2), CS ($J$=7--6), and the SO ($J_N$=7$_8$--6$_7$) lines
with the Atacama Submillimeter Telescope Experiment (ASTE) 10-m telescope.
The ASTE mapping data in the C$^{18}$O line are combined
with our previous SMA data in the same line, which have a larger field of view and a shorter spacing
than those of the ALMA data and thus more suitable to study the
connection between the extended protostellar envelope and the CBD.
In this paper we report these ASTE+SMA results of the protostellar envelope in L1551 NE.
The structure of this paper is as follows. In section 2,
our new ASTE observations, as well as the process to combine the ASTE mapping data
with the SMA data in the C$^{18}$O line, are described. In section 3,
the results of the ASTE mapping observations and the combination of the ASTE + SMA data
are presented. In section 4, gas motions and physical properties of the identified gas components
are estimated. In section 5, we discuss implications of the observed features in the context of
growth of protostellar binaries, and concise summary is written in section 6.

\section{ASTE Observations}

We performed observations of L1551 NE in the C$^{18}$O ($J$=3--2; 329.3305453 GHz),
$^{13}$CO ($J$=3--2; 330.587965 GHz), SO ($J_N$=7$_8$--6$_7$; 340.71416 GHz), and
the CS ($J$=7--6; 342.882857 GHz) lines with the ASTE 10-m telescope
on 2013 September 23-26 and October 4.
Remote observations were performed from the ASTE operation room of NAOJ at Mitaka,
Japan, using the network observation system N-COSMOS3 developed by NAOJ \cite{kam05}.
A cartridge-type, double-sideband 350 GHz receiver with an IF frequency range of 4.5 to 7 GHz (CATS345)
mounted on ASTE \cite{koh05} was used, and all the four lines were observed simultaneously,
except that on September 23 only the SO and CS lines were observed due to the mistake
of the instrumental setting.
We used only the data with the
DSB system noise temperature ranging $\sim$200 - 800 K,
and the data with the higher noise temperatures are excluded.
The telescope pointing was checked every
$\sim$1.5 - 2 hours by five-point CO ($J$ = 3--2) observations of
a late-type star, NML-Tau, and was found to be better than $\sim$2$\arcsec$.
As a standard source we also observed Orion KL \cite{sch97} and L1551 IRS 5 \cite{tak11},
and confirmed that the relative intensity was consistent within $\sim$30$\%$
with a main beam efficiency of 0.6.
No further correction for the sideband ratio was performed.
Hereafter we show the observed line intensities in the unit of $T_{MB}$.


The observations consisted of two parts; one-point, deep observations
toward the center of L1551 NE ($\alpha_{J2000}$,$\delta_{J2000}$)=
(04$^{h}$31$^{m}$44$\fs$47, 18\arcdeg08\arcmin32$\farcs$2),
which is the field center of our previous SMA observations and matches approximately
the positions of the protostellar binary \cite{tak12,tak13},
and mapping observations around L1551 NE.
Both observations were conducted in the position-switching mode.
The SO and CS line data toward the center of L1551 NE taken on September 23,
when the other line data were not taken due to the mistake of the instrumental setting,
are included in the deep integrations.
The resultant total on-source integration times of the SO and CS data toward
the center of L1551 NE is 1620 sec, while those of the $^{13}$CO and C$^{18}$O
data 1220 sec. The rms noise levels of the $^{13}$CO, C$^{18}$O, SO, and
the CS spectra toward the center of L1551 NE are 0.055 K, 0.062 K, 0.025 K, and
0.029 K, respectively.
The mapping observations were conducted with a grid spacing of 10$\arcsec$ (Nyquist sampling),
and a typical on-source integration time per point of $\sim$80 seconds.
The map center is set to be the position of the deep integrations.
The spatial and spectral resolutions and the noise level of the mapping observations in the
C$^{18}$O line are summarized in Table 1.

We combined the ASTE image cube in the C$^{18}$O line with the interferometric C$^{18}$O data
of L1551 NE taken with the SMA in its subcompact and compact configurations \cite{tak13},
adopting the method described by Takakuwa et al. (2007b).
Details of the SMA observations are described by Takakuwa et al. (2013).
The conversion factor from $T_{MB}$ (K) to S (Jy beam$^{-1}$) was derived to be 48.5 as
\begin{equation} 
S = \frac{2k_{B}\Omega_{beam}}{\lambda^2}T_{MB},
\end{equation}
where $k_{B}$ is the Boltzmann constant, $\lambda$ is the wavelength,
and $\Omega_{beam}$ is the solid angle of the ASTE beam (= 23$\arcsec$).
The spatial and spectral resolutions and the noise level of the combined SMA+ASTE image cube in the
C$^{18}$O line are summarized in Table 1.

\section{Results}
\subsection{ASTE Spectra toward L1551 NE}

\begin{figure}
\epsscale{0.8}
\plotone{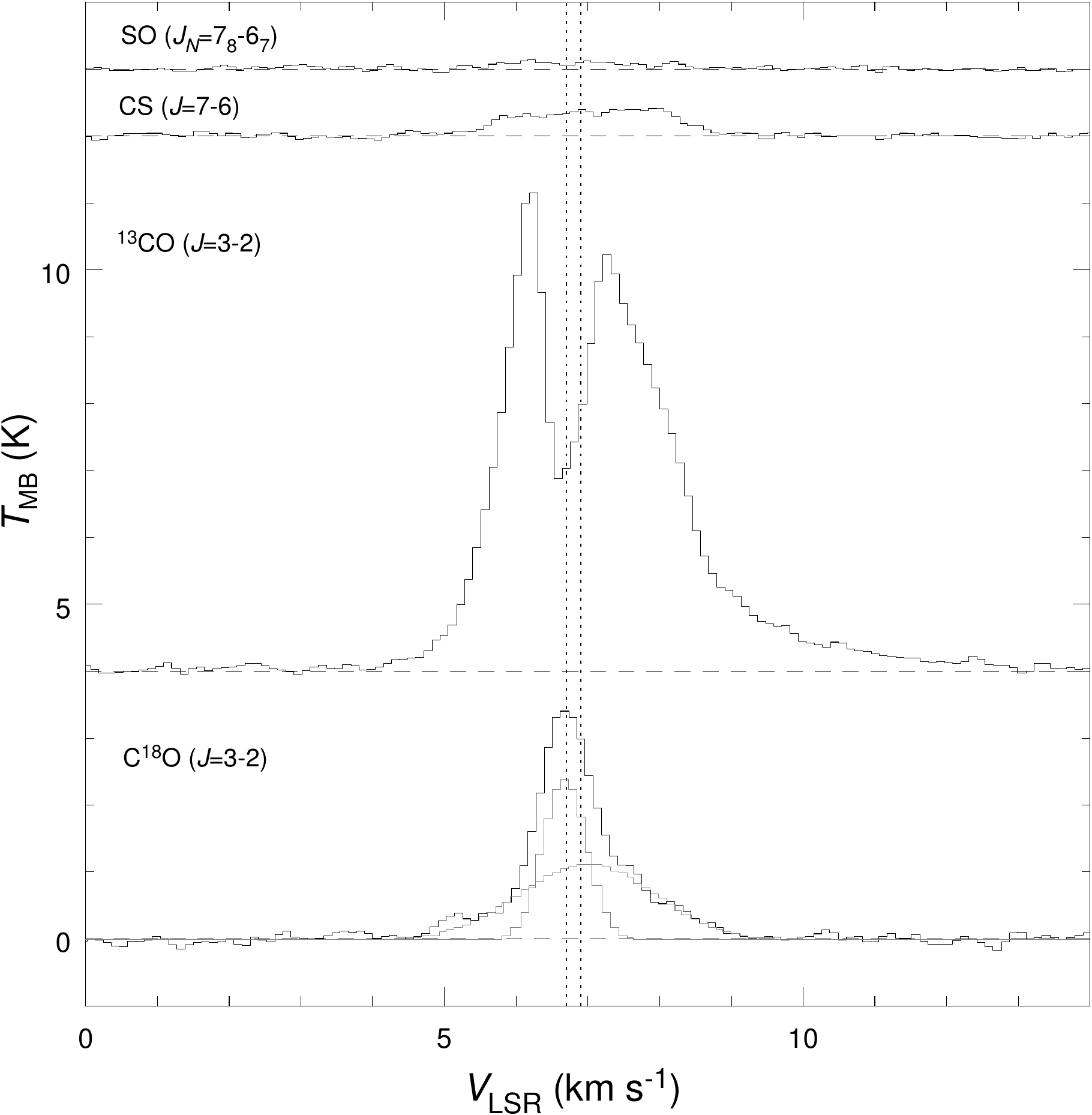}
\caption{ASTE spectra toward the central position of L1551 NE.
Light lines at the bottom show the results of the two-component Gaussian fitting to the
observed C$^{18}$O (3--2) spectrum. The vertical dashed lines denote
the centroid velocities of the envelope ($V_{LSR}$ = 6.7 km s$^{-1}$) and
the central circumbinary disk ($V_{LSR}$ = 6.9 km s$^{-1}$).
\label{astesp}}
\end{figure}

Figure \ref{astesp} shows the observed ASTE spectra of the
C$^{18}$O ($J$=3--2), $^{13}$CO ($J$=3--2), CS ($J$=7--6), and the
SO ($J_N$=7$_8$--6$_7$) lines toward the protobinary of L1551 NE.
The C$^{18}$O spectrum appears to consist of narrow and wide components.
Two-component Gaussian fitting to the C$^{18}$O spectrum
shows that the peak brightness temperature, line width, and the centroid velocity
%
of the narrow component are 2.4 K, 0.72 km s$^{-1}$, and 6.68 km s$^{-1}$,
and those of the wide component 1.1 K, 2.24 km s$^{-1}$, and 6.99 km s$^{-1}$,
respectively (light lines in Figure \ref{astesp}).
Previous CSO observations
also found that the C$^{18}$O (3--2) spectrum toward L1551 NE
can be decomposed of two Gaussian components with similar line widths
and centroid velocities \cite{ful02}, 
In particular, the centroid velocity of the broad component is similar to the centroid
velocity of the CBD derived from our previous SMA observations (= 6.9 km s$^{-1}$)
\cite{tak12,tak13}. As shown in the next subsection, the broad C$^{18}$O component is
present only toward the protobinary position.
These results indicate that the broader component in the ASTE C$^{18}$O spectrum
likely originates from the central CBD. On the other hand,
the narrower C$^{18}$O component traces a distinct component
at a slightly blueshifted ($\sim$0.2 km s$^{-1}$) centroid velocity.
As shown below, this narrow component traces the extended envelope component
around L1551 NE.
Hereafter in this paper, $V_{LSR}$ = 6.7 km s$^{-1}$ is adopted as the systemic
velocity of the extended envelope component, while $V_{LSR}$ = 6.9 km s$^{-1}$
as the systemic velocity of the CBD in L1551 NE.

The $^{13}$CO ($J$=3--2) spectrum shows an absorption dip at around
the centroid velocity of the narrow component of the C$^{18}$O spectrum,
and a broad redshifted wing up to $V_{LSR}$ $\sim$13 km s$^{-1}$.
While a part of
the redshifted wing in the $^{13}$CO spectrum likely originates from the
CBD as in the case of the C$^{18}$O spectrum, mapping
observations of the $^{13}$CO line demonstrate that the redshifted $^{13}$CO emission
is most likely arising from the outflow component driven from L1551 IRS 5,
as will be discussed below.
With a deeper integration a much better CS ($J$=7--6) spectrum of L1551 NE
than our previous one \cite{tak11} is obtained.
The CS spectrum shows a non-Gaussian, flat-top spectral shape. 
Whereas the SO ($J_N$=7$_8$--6$_7$) line is weak compared to the other lines,
the peak intensity ($\sim$0.14 K) is detected above 5.5$\sigma$.

\subsection{Spatial and Velocity Distributions Observed with ASTE}

\begin{figure}
\epsscale{1.0}
\plotone{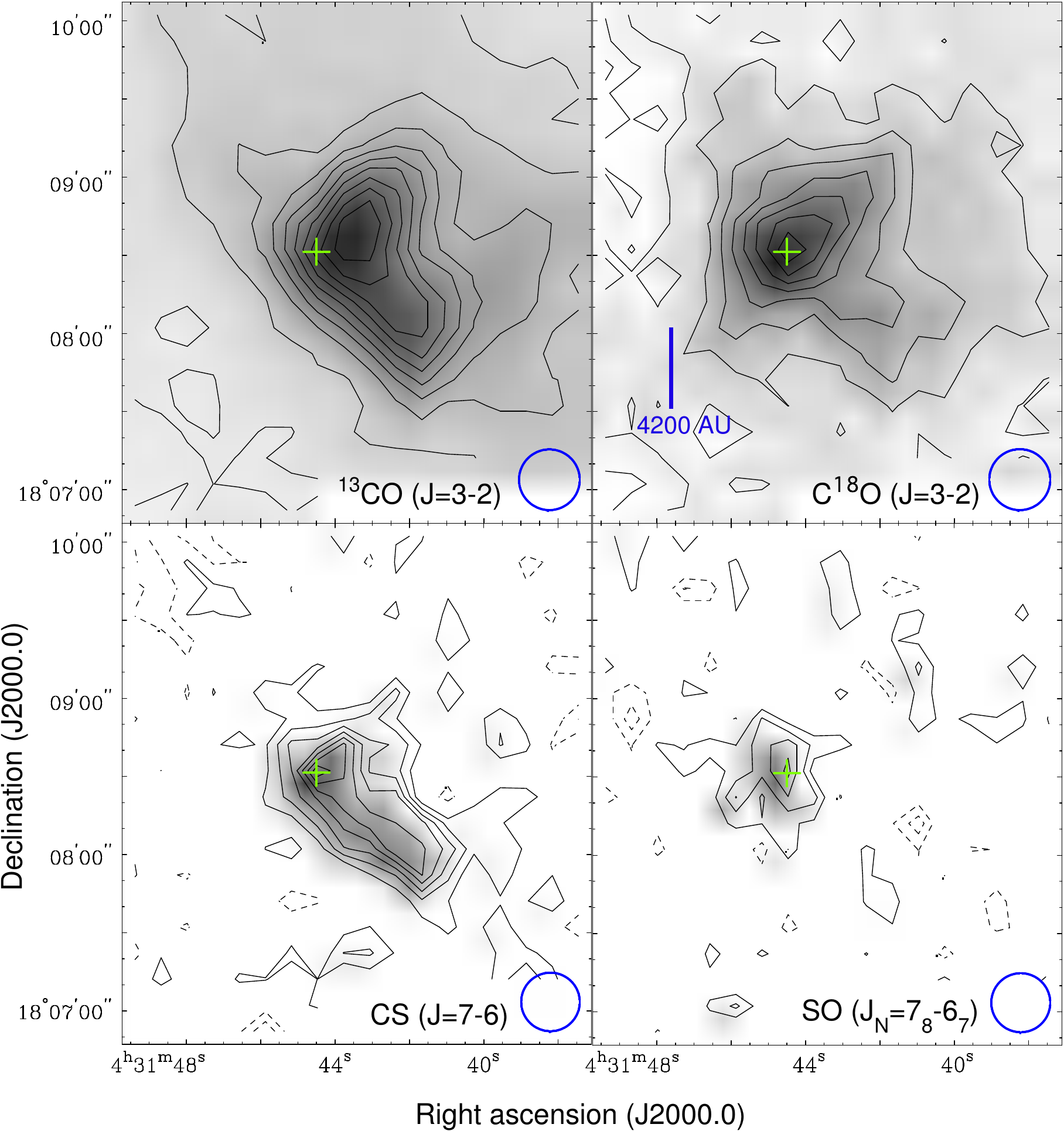}
\caption{ASTE total integrated intensity maps of the observed lines
as labeled. Crosses show the position of L1551 NE.
Integrated velocity ranges are 4.48 -- 12.53 km s$^{-1}$,
5.31 -- 9.86 km s$^{-1}$, 5.63 -- 9.13 km s$^{-1}$, and
5.69 -- 8.00 km s$^{-1}$, for the $^{13}$CO (3--2), C$^{18}$O (3--2),
CS (7--6), and the SO (7$_{8}$--6$_{7}$) maps, respectively.
Contour levels are 15$\sigma$, 20$\sigma$, and then in steps of
15$\sigma$ (1$\sigma$ = 0.137 K km s$^{-1}$) for the $^{13}$CO (3--2) map,
2$\sigma$, 5$\sigma$, and then in steps of 5$\sigma$
(1$\sigma$ = 0.130 K km s$^{-1}$) for the C$^{18}$O (3--2) map,
and in steps of 2$\sigma$ for the CS (7--6) and SO (7$_{8}$--6$_{7}$) maps
(1$\sigma$ = 0.0633 K km s$^{-1}$ and 0.0517 K km s$^{-1}$ for the CS and SO maps,
respectively.).
Open circles at the bottom-right corners denote the ASTE beam of the relevant lines.
\label{astemom0}}
\end{figure}

Figure \ref{astemom0} shows total integrated intensity maps of the four molecular lines
observed with ASTE. The C$^{18}$O (3--2) emission
exhibits condensed and diffuse components, the former of which is centered
on the protobinary position.
Two-component, two-dimensional Gaussian fitting
to the C$^{18}$O total integrated map shows that the C$^{18}$O emission can be decomposed
of a central component with a size of $\sim$5500 AU $\times$ 2600 AU (P.A.=-46$\degr$)
and a diffuse component with a size of $\sim$21000 AU.
Emission peaks toward the protostellar position are also seen in the CS (7--6)
and SO (7$_8$--6$_7$) lines.
On the other hand, the $^{13}$CO (3--2) peak emission
does not coincide with the protostellar position but is offset at $\sim$20$\arcsec$ west from the protostar.
The $^{13}$CO emission shows an elongated feature toward the southwest
direction, which is also evident in the C$^{18}$O and CS emission.

\begin{figure}
\epsscale{0.95}
\plotone{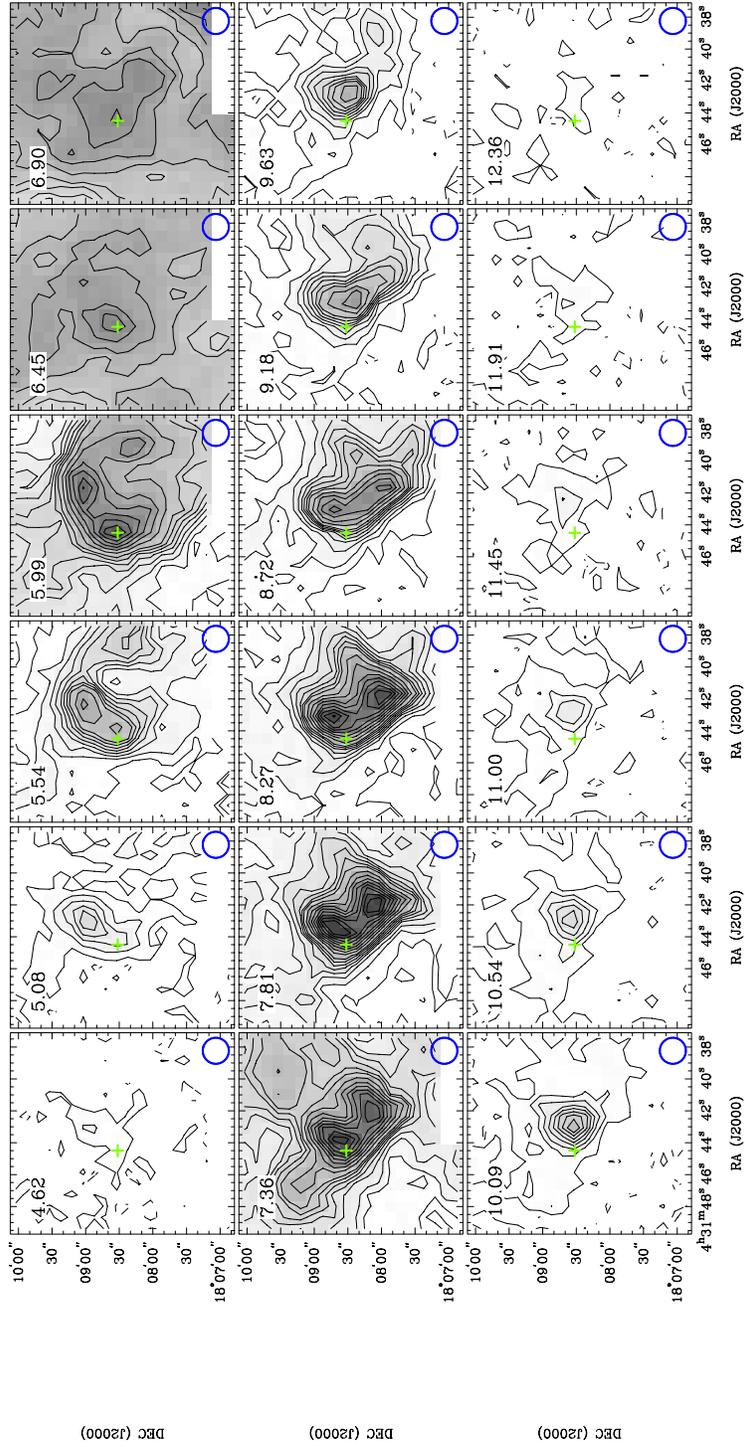}
\caption{ASTE velocity channel maps of the $^{13}$CO (3--2) line in L1551 NE at a velocity
bin of 0.455 km s$^{-1}$.
Contour levels start from 2$\sigma$ in steps of 4$\sigma$ until 30$\sigma$, and then
in steps of 8$\sigma$ (1$\sigma$ = 0.0717 K). Symbols are the same
as those in Figure \ref{astemom0}.
\label{aste13co}}
\end{figure}

Figure \ref{aste13co} shows the ASTE velocity channel maps of the $^{13}$CO line
at a velocity interval of 0.455 km s$^{-1}$. In the blueshifted velocity range
($V_{LSR}$=4.62 -- 5.99 km s$^{-1}$), an emission component to the northwest
of the protostar and that centered on the protostellar position are seen.
Around the systemic velocity (6.45 -- 6.90 km s$^{-1}$) the $^{13}$CO line
shows less intensity contrast, suggesting that the $^{13}$CO emission traces
the overall cloud component. In the redshifted velocity of 7.36 -- 9.18 km s$^{-1}$,
two emission components, one located to $\sim$20$\arcsec$ west of the protostar,
and the other to $\sim$40$\arcsec$ southwest, are seen. While the southwestern
component diminishes at velocities higher than 9.63 km s$^{-1}$, the western component
is seen until $\sim$12.36 km s$^{-1}$. The location and velocity of this
high-velocity redshifted component is consistent with those of a
redshifted outflow component seen in the $^{12}$CO (1--0; 2--1; 3--2) lines \cite{sto06,mor06}
and the CS (2--1; 3--2) lines \cite{pla95,yok03} driven
from the neighboring protostellar source, L1551 IRS 5.
Thus, the high-velocity redshifted component seen in the $^{13}$CO line most likely
traces the redshifted outflow component from L1551 IRS 5.

\begin{figure}
\epsscale{0.95}
\plotone{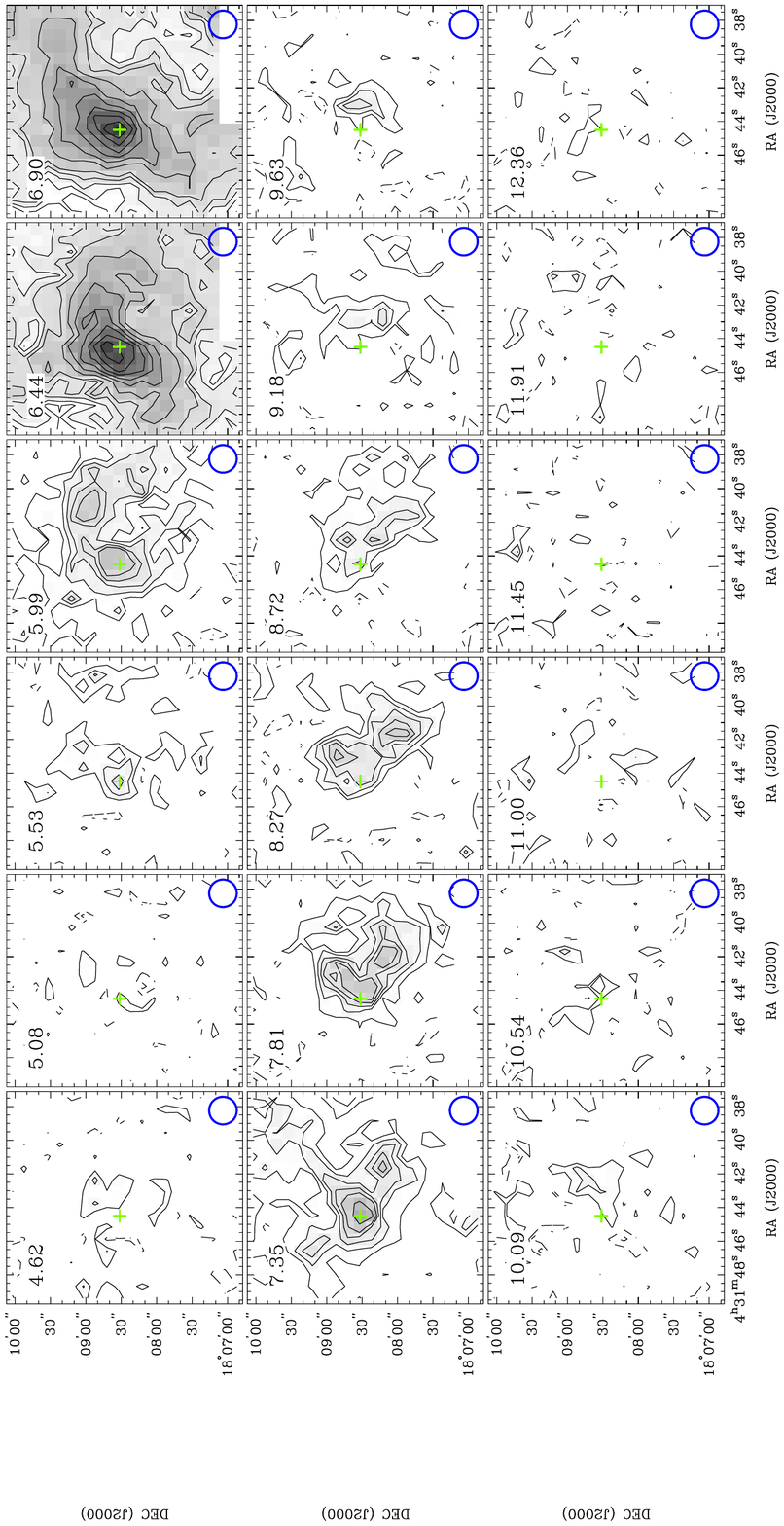}
\caption{Same as Figure \ref{aste13co} but for the C$^{18}$O (3--2) line.
Contour levels start from 2$\sigma$ in steps of 2$\sigma$ until 10$\sigma$, and then
in steps of 4$\sigma$ (1$\sigma$ = 0.090 K). \label{astec18o}}
\end{figure}

Figure \ref{astec18o} shows the ASTE velocity channel maps of the C$^{18}$O line
at the same velocity bin as that of the $^{13}$CO velocity channel maps.
The C$^{18}$O emission component associated with the protostar is seen from
$V_{LSR}$=5.53 km s$^{-1}$. The C$^{18}$O emission component
exhibits an elongated feature at 6.44 -- 6.90 km s$^{-1}$ along the northwest to
the southeast direction, which is approximately consistent with the major axis
of the central disk. While the C$^{18}$O emission peak at 6.44 km s$^{-1}$ is
located to the northeast of the protostar, the emission location is systematically shifted
from northeast to southwest from the blueshifted to redshifted velocities
(6.44 km s$^{-1}$ to 7.81 km s$^{-1}$). These results indicate that there is a velocity
gradient in the C$^{18}$O emission along the direction of the minor axis of the central disk.
These C$^{18}$O emission components most likely trace the protostellar envelope
surrounding L1551 NE (hereafter ``ENV'').
In the redshifted velocities from 7.35 km s$^{-1}$, another emission component
to the southwest of the protostar becomes evident. This component appears to trace the same
southwestern gas component seen in the $^{13}$CO emission. In even higher redshifted velocities
(8.72 -- 10.54 km s$^{-1}$), the C$^{18}$O counterpart of the western redshifted component
found in the $^{13}$CO emission is also seen, although the highest velocity end of the
C$^{18}$O counterpart is lower than that of the $^{13}$CO emission.

\begin{figure}
\epsscale{0.95}
\plotone{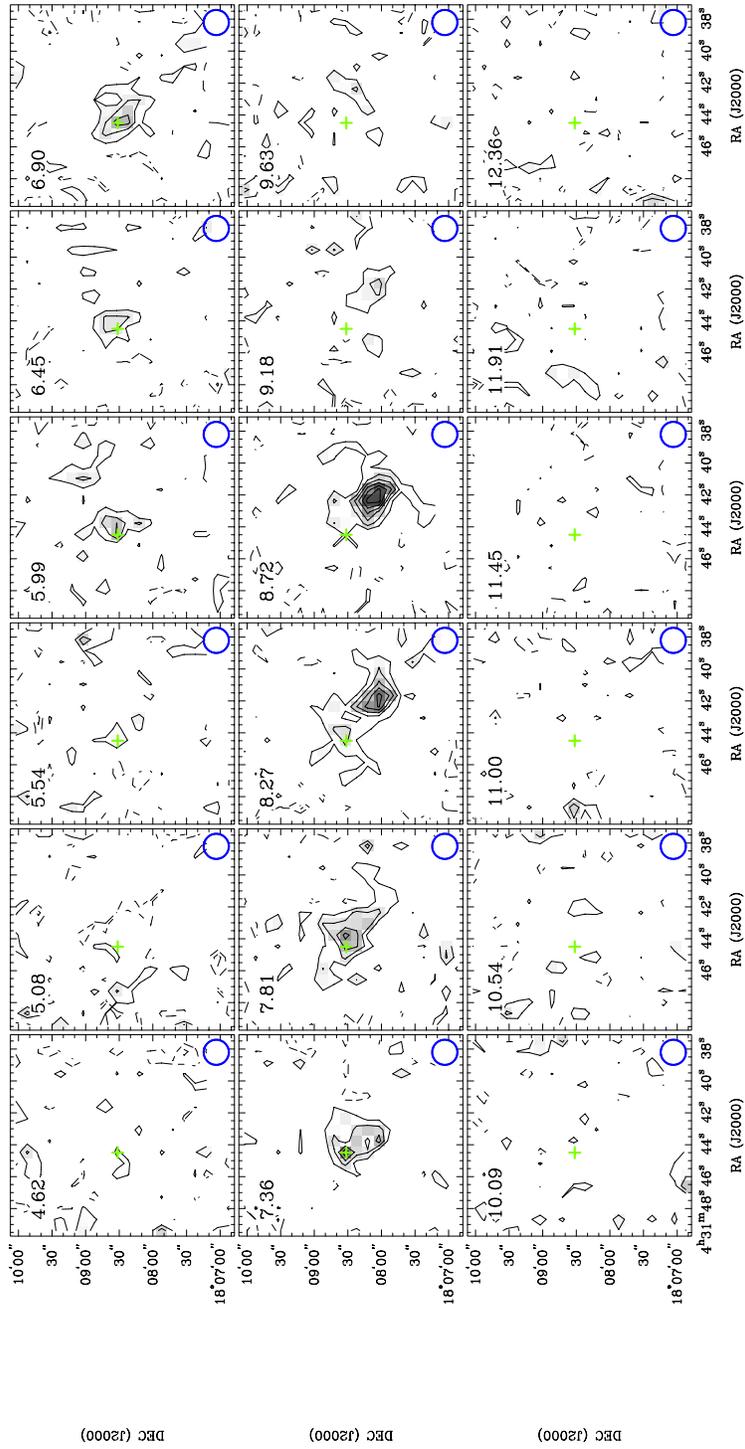}
\caption{Same as Figure \ref{aste13co} but for the CS (7--6) line.
Contour levels start from 2$\sigma$ in steps of 2$\sigma$ (1$\sigma$ = 0.050 K).
\label{astecs}}
\end{figure}

Figure \ref{astecs} shows the same velocity channel maps as those of
Figures \ref{aste13co} and \ref{astec18o} but for the CS (7--6) line.
While the CS emission traces a gas component associated with the protostar
in the velocity range of 5.54 -- 7.81 km s$^{-1}$, the most prominent CS emission
component is the redshifted (8.27 -- 9.18 km s$^{-1}$) component to the southwest of the protostar.
This southwestern redshifted component is also seen in the $^{13}$CO and C$^{18}$O emission.
Previous BIMA and NMA observations of the CS (2--1; 3--2) lines have also found the same
CS component, where the CS abundance is enhanced by a factor of a few \cite{pla95,yok03}.
This CS component is considered to be shock-excited
molecular gas of the outflow driven from L1551 IRS 5 \cite{pla95}. The
detection of the submillimeter CS (7--6) emission with the present ASTE observations,
which traces warm ($\gtrsim$ 60 K) molecular gas \cite{ta07a,tak11}, further supports this interpretation.
Hereafter we call this redshifted component to the southwest of the protostar ``RED1'',
and the highest-velocity redshifted component seen in the $^{13}$CO and C$^{18}$O lines
to the west of the protostar ``RED2''.

\begin{figure}
\epsscale{1.0}
\plotone{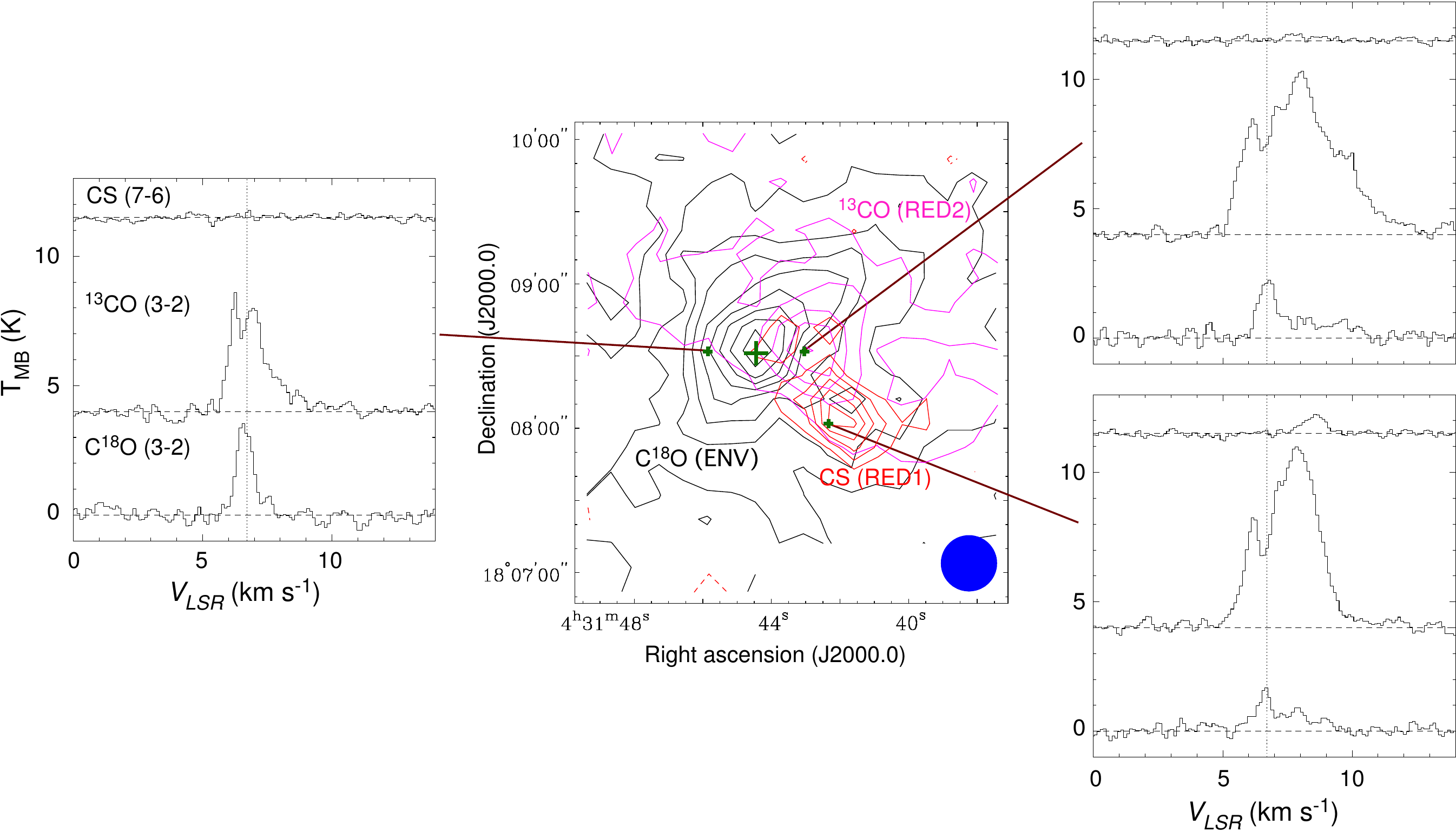}
\caption{Maps of the envelope component in the C$^{18}$O (3--2) line (black contours),
the high-velocity redshifted component in the $^{13}$CO (3--2) line (purple contours),
and the redshifted CS (7--6) component (red contours), along with the line profiles
toward the representative positions marked with crosses. Velocity ranges of the
C$^{18}$O, $^{13}$CO, and the CS maps are 5.31 - 8.15 km s$^{-1}$,
9.47 - 12.53 km s$^{-1}$, and 8.04 - 9.13 km s$^{-1}$, respectively.
Contour levels start from 5$\sigma$ in steps of 5$\sigma$ in the C$^{18}$O map (1$\sigma$ = 0.036 K),
5$\sigma$ in steps of 10$\sigma$ in the $^{13}$CO map (1$\sigma$ = 0.028 K),
and 4$\sigma$ in steps of 3$\sigma$ in the CS map (1$\sigma$ = 0.032 K).
A large cross and a filled circle at the bottom-right corner denote the protostellar position
and the ASTE beam in the C$^{18}$O map, respectively. Vertical dashed lines in the spectra
show the systemic velocity of the envelope of $V_{LSR}$ = 6.7 km s$^{-1}$.
\label{colli}}
\end{figure}

In Figure \ref{colli}, we compare spatial distributions of ENV as seen in the C$^{18}$O emission,
RED1 in the CS emission, and RED2 in the $^{13}$CO emission. Line profiles in the peak positions
of RED1 and RED2, as well as those 20$\arcsec$ east of the protostar, are also shown
(Note that the line profiles of ENV are shown in Figure \ref{astesp}.). The two outflow components
of RED1 and RED2 are located to the western side of ENV.
Toward RED1, the $^{13}$CO line profile shows an intense redshifted peak plus a red wing
up to $V_{LSR}$ $\sim$12 km s$^{-1}$. The C$^{18}$O line profile exhibits redshifted emission ``plateau''
up to $V_{LSR}$ $\sim$10 km s$^{-1}$, and the CS line profile shows the emission only in the redshifted
velocity range. Similar $^{13}$CO redshifted wing and C$^{18}$O emission plateau are also seen
toward RED2, where a weak redshifted CS emission is also present.
On the other hand, toward the 20$\arcsec$ east of the protostar, such a C$^{18}$O emission plateau
is not seen and the $^{13}$CO redshifted wing is much less prominent, and the submillimeter
CS emission is not detected.
These results imply that the outflow streams of RED1 and RED2 are terminated
at the western side of ENV. One possible interpretation for this termination is
the collision between the outflow components and ENV.

\begin{figure}
\epsscale{1.0}
\plotone{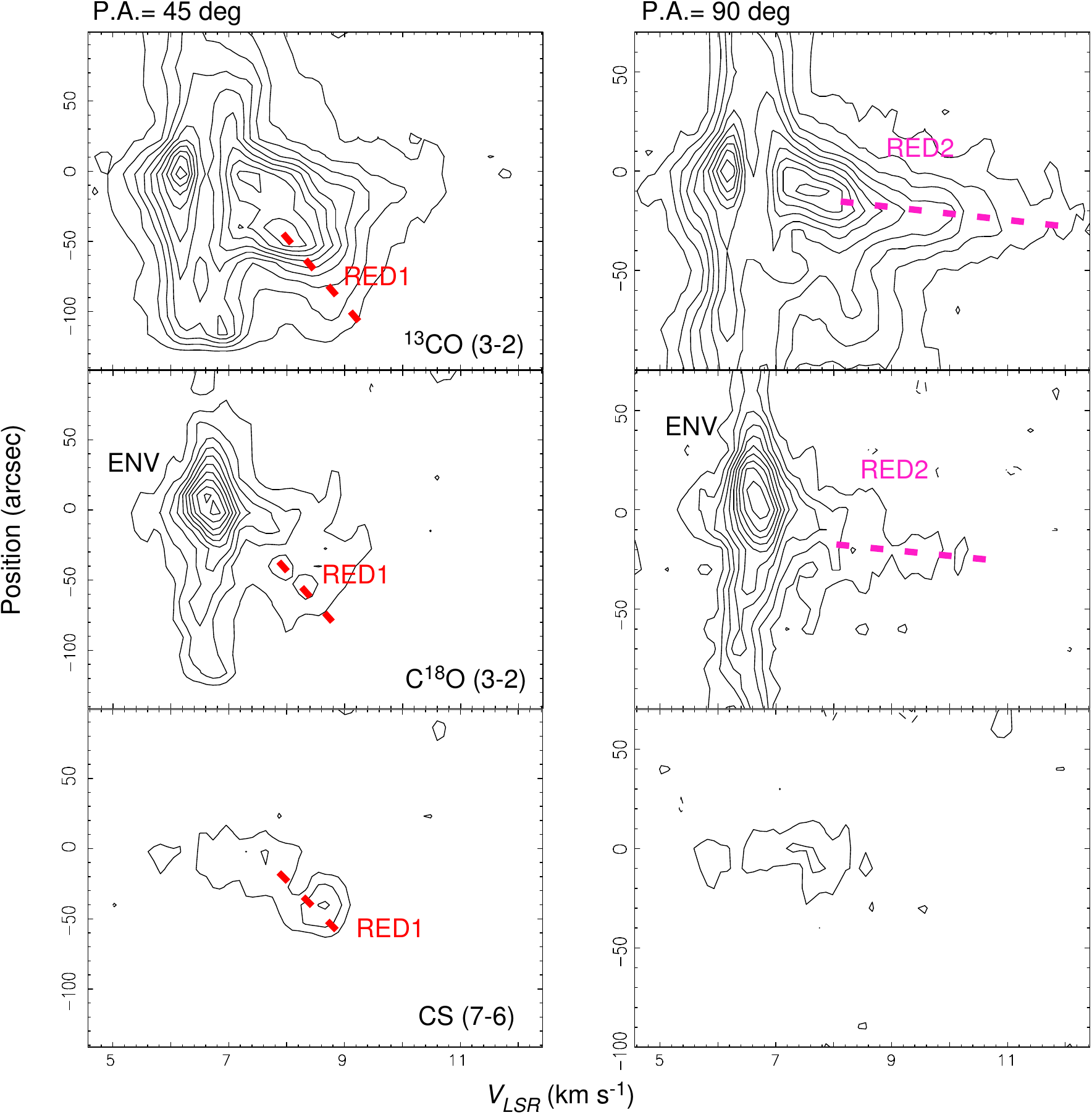}
\caption{ASTE Position-Velocity (P-V) diagrams of the $^{13}$CO (3--2) (upper panels),
C$^{18}$O (3--2) (middle), and the CS (7--6) lines (lower) along the cuts passing through
the protostellar position at P.A. = 45$\degr$ (left panels) and P.A. = 90$\degr$ (right).
These cut lines
connect the protostellar position and the positions of RED1 and RED2 shown in
Figure \ref{colli}. Contour levels start from 2.5$\sigma$ in steps of 5$\sigma$
in the $^{13}$CO P-Vs (1$\sigma$ = 0.143 K), and in steps of 2$\sigma$ for
the C$^{18}$O (1$\sigma$ = 0.18 K) and CS P-Vs (1$\sigma$ = 0.1 K). For direct comparison
the velocity bins of the three P-Vs are aligned to be the same with the width of 0.114 km s$^{-1}$.
Red and purple dashed lines delineate the detected velocity gradients of RED1 and RED2, respectively.
\label{collipv}}
\end{figure}

In Figure \ref{collipv}, we show the Position-Velocity (P-V) diagrams
of the $^{13}$CO, C$^{18}$O, and the CS lines along the cuts passing through the protostellar position
and the peak positions of RED1 (P.A. = 45$\degr$) and RED2 (P.A. = 90$\degr$).
In the C$^{18}$O P-Vs the envelope component is seen as labeled. To the west of ENV
the $^{13}$CO, C$^{18}$O, and the CS emission components arising from RED1,
and the $^{13}$CO and C$^{18}$O emission components from RED2, can be identified.
There appears velocity gradients both in RED1 and RED2 as delineated
by dashed lines, and the velocities of RED1 and RED2 become closer to the systemic velocity
as the positions become closer to ENV. These results suggest that the flow velocities
of RED1 and RED2 are decelerated as they become close to ENV, presumably due to
the collision with ENV.

\begin{figure}
\epsscale{1.0}
\plotone{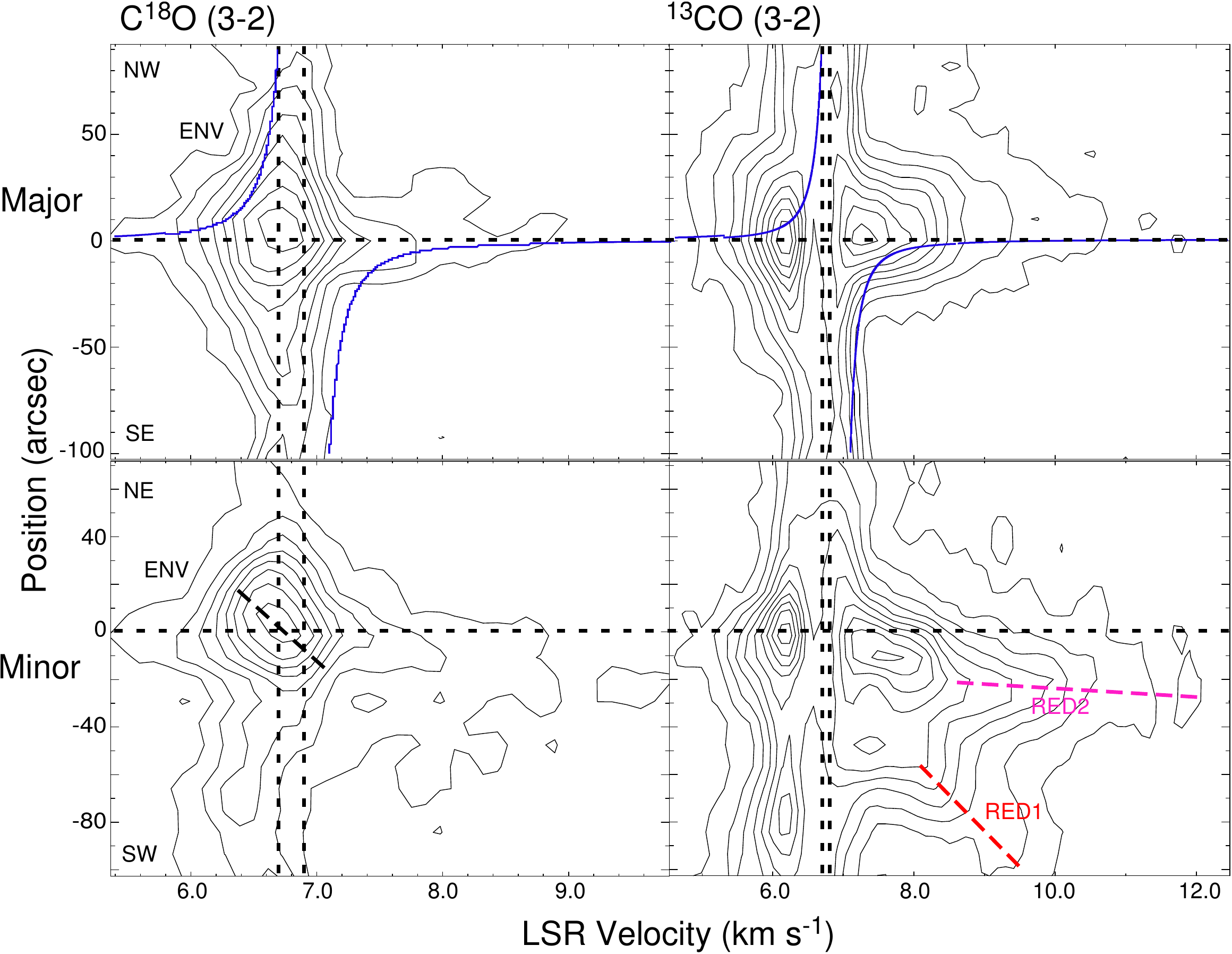}
\caption{ASTE P-V diagrams of the C$^{18}$O (3--2) (left panels)
and $^{13}$CO (3--2) (right) lines along the major (P.A.= 167$\degr$; upper panels) and
minor (P.A.= 77$\degr$; lower) axes of the Keplerian CBD around L1551 NE.
Contour levels are in steps of 2.5$\sigma$ for the C$^{18}$O P-Vs (1$\sigma$ = 0.180 K),
and start from 2.5$\sigma$ in steps of 5$\sigma$ for the $^{13}$CO P-Vs (1$\sigma$ = 0.143 K).
Horizontal dashed lines show the protostellar position, and the vertical dashed lines the centroid
velocities of the extended C$^{18}$O envelope (= 6.7 km s$^{-1}$)
and the Keplerian CBD (= 6.9 km s$^{-1}$). Blue curves show
the rotation curve of the central Keplerian CBD.
A black tilted dashed line delineates the detected velocity gradient of the envelope,
and red and purple dashed lines that of RED1 and RED2, respectively.
\label{astepvs}}
\end{figure}

Figure \ref{astepvs} shows ASTE P-V diagrams of the C$^{18}$O and $^{13}$CO lines
along the major and minor axes of the central CBD passing through the
protostellar position. In the C$^{18}$O P-Vs, the line broadening at the protostellar position
arising from the CBD is clearly seen, and the velocity gradient originated
from the Keplerian rotation (blue curves in Figure \ref{astepvs}) can be identified along the major axis.
As discussed in the C$^{18}$O line profile toward the protostellar position
(Figure \ref{astesp}), the centroid velocity of the CBD component and that of the envelope component
are offset by $\sim$0.2 km s$^{-1}$ (two vertical dashed lines in Figure \ref{astepvs}),
and the C$^{18}$O P-Vs exhibit this offset clearly. Furthermore, along the major axis
the envelope component as seen in the C$^{18}$O emission does not show any noticeable velocity gradient,
and the Keplerian rotation of the central disk
is not continuous to the outer envelope. On the other hand,
along the minor axis the C$^{18}$O envelope shows a clear velocity gradient
along the northeast (blue) to southwest (red) direction (tilted black
dashed line in Figure \ref{astepvs}). Since the associated NIR jets are blueshifted to the southwest
and redshifted to the northeast \cite{hay09}, the southwestern side of the disk and envelope
must be far-side and the northeastern side near-side, assuming that the midplane of the disk
and envelope is perpendicular to the axis of the jets.
Correspondingly, the southwestern redshifted component 
would then be located on the far side and the northeastern blueshifted component
on the near side, suggesting the outward radial motion on the plane.
In the $^{13}$CO P-V diagram along the minor axis, both RED1 and RED2
and their decelerated velocity structures (red and pink dashed lines) are identified.
These results indicate that the envelope is being dissipated due to the interaction
with the redshifted outflowing gas driven from L1551 IRS 5.
We will further discuss this dispersing motion of the envelope with the combined SMA+ASTE images
of the C$^{18}$O emission below.

\subsection{Spatial and Velocity Structures of
the Inner Envelope and\\
Circumbinary Disk traced by the SMA+ASTE C$^{18}$O (3--2) Image Cube}

\begin{figure}
\epsscale{1.0}
\plotone{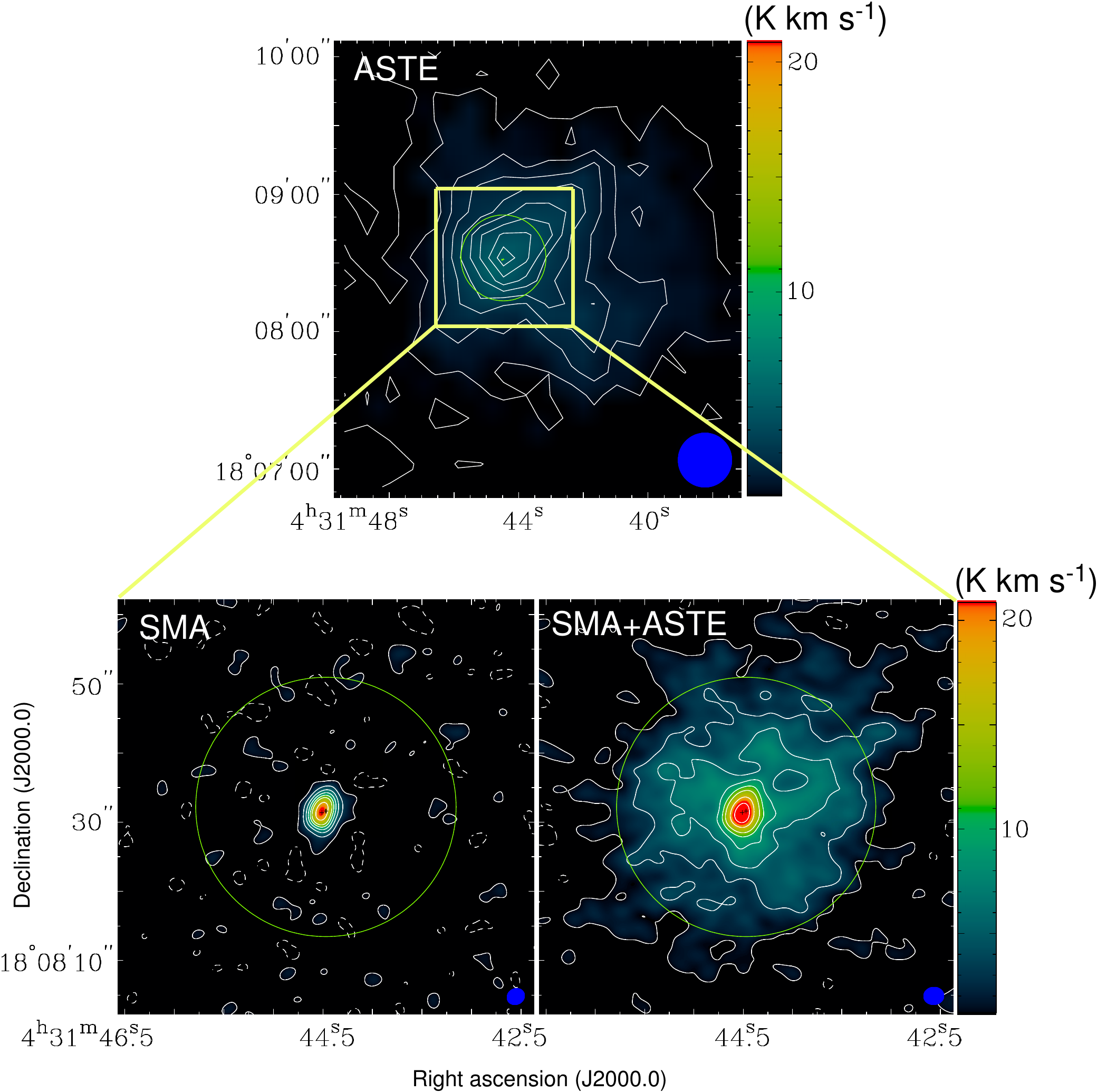}
\caption{ASTE, SMA, and SMA+ASTE total integrated intensity maps
of the C$^{18}$O (3--2) line in L1551 NE as labeled. Integrated velocity
ranges of the ASTE, SMA, and SMA+ASTE maps are
5.31 -- 9.86 km s$^{-1}$, 3.75 -- 9.49 km s$^{-1}$, and 3.94 -- 9.30 km s$^{-1}$,
respectively. An rms noise level of the SMA+ASTE map is
$\sigma_{comb}$ = 0.596 K km s$^{-1}$, and the contour levels of the SMA
and SMA+ASTE maps start from 2$\sigma_{comb}$ in steps of
4$\sigma_{comb}$
until 22$\sigma_{comb}$, and then in steps of 8$\sigma_{comb}$.
Contour levels of the ASTE map are in steps of 1$\sigma_{comb}$.
Rms noise levels of the ASTE and SMA maps are 1$\sigma_{\rm ASTE}$ = 0.130 K km s$^{-1}$
and 1$\sigma_{\rm SMA}$ = 0.379 K km s$^{-1}$, respectively.
Open circles and crosses show the field of view of the SMA and the
positions of the protostellar binary, respectively. Filled ellipses at the bottom-right corners denote
the beam sizes (23$\farcs$4 in the ASTE map,
2$\farcs$50 $\times$2$\farcs$20; P.A. = -60$\fdg$6 in the SMA map, and 
2$\farcs$85 $\times$2$\farcs$45; P.A. = -85$\fdg$3 in the SMA+ASTE map).
\label{mom0s}}
\end{figure}

Figure \ref{mom0s} compares moment 0 maps of the ASTE, SMA, and the combined SMA+ASTE
image cubes of the C$^{18}$O (3--2) line in L1551 NE. While the ASTE image exhibits an extended
envelope structure as described in the last subsection,
the SMA image shows the central smaller-scale ($\sim$1000 AU $\times$ 800 AU) structure
with its position angle ($\sim$167$\degr$) consistent with that of the Keplerian CBD
\cite{tak13}. Combining these
two images shows that the central structure found by the SMA observations is embedded
in the more extended envelope.

\begin{figure}
\epsscale{0.95}
\plotone{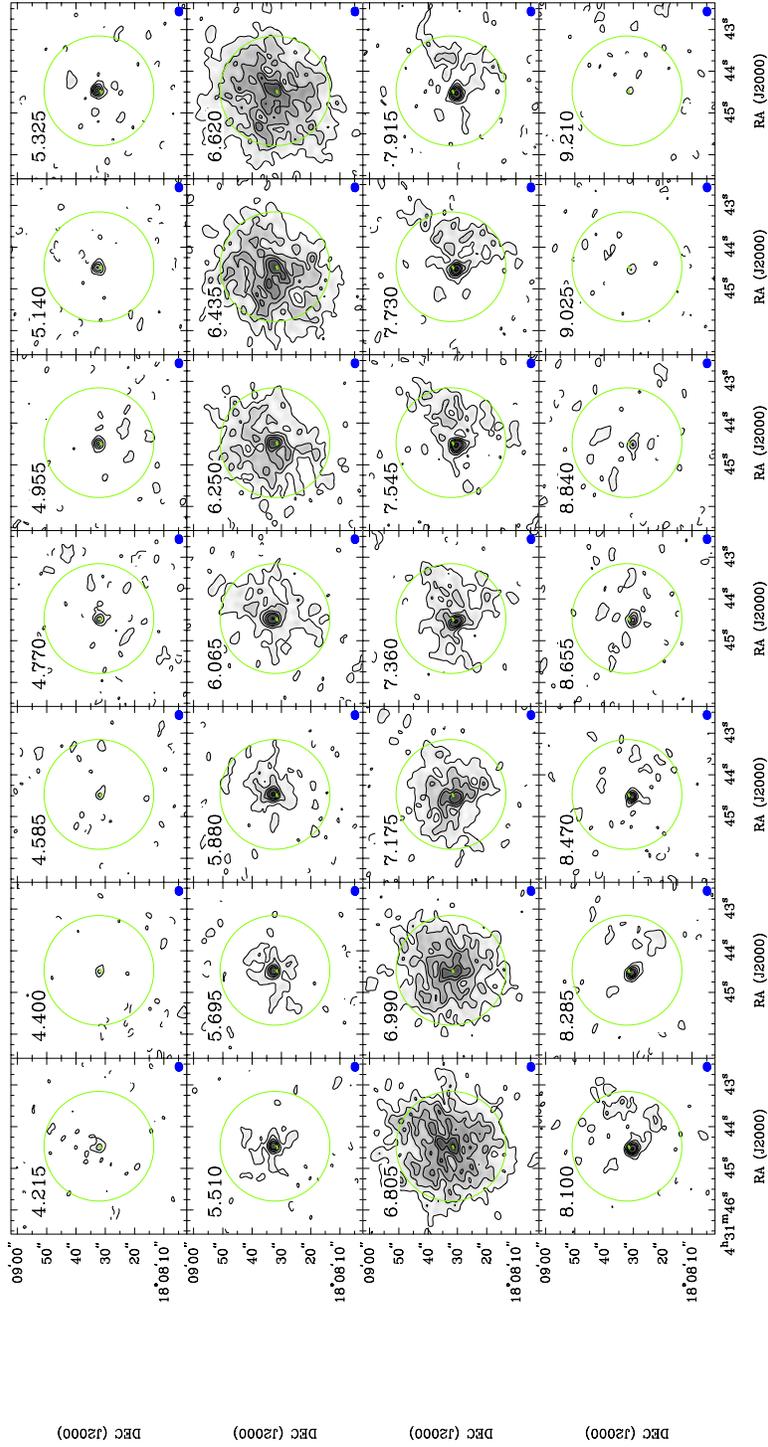}
\caption{SMA+ASTE velocity channel maps of the C$^{18}$O (3--2) line in L1551 NE.
Contour levels start from 2$\sigma$ in steps of 2$\sigma$ until 10$\sigma$, and then
in steps of 4$\sigma$ (1$\sigma$ = 0.598 K). Symbols are the same
as those in Figure \ref{mom0s}.
\label{combch}}
\end{figure}

Figure \ref{combch} shows velocity channel maps of the SMA+ASTE image cube
of the C$^{18}$O (3--2) line. In the highly blueshifted ($V_{LSR}$ = 4.2 -- 6.1 km s$^{-1}$)
and redshifted (7.5 -- 9.2 km s$^{-1}$)
velocities, compact C$^{18}$O emission located to the north and south of the protostellar binary are seen,
respectively.
These compact components trace the Keplerian CBD
as we already discussed in our previous papers \cite{tak12,tak13,tak14}.
The extended envelope component appears in the lower velocity range
(6.1 -- 8.2 km s$^{-1}$). In the blueshifted velocity range of 6.3 -- 6.6 km s$^{-1}$,
the extended C$^{18}$O emission is located predominantly to the eastern side of
the protostellar binary, whereas in the redshifted velocity range of 7.5 -- 8.2 km s$^{-1}$
to the western side.
Thus, there is a velocity gradient along the east (blue) to west
(red) direction in the envelope, which is also identified in the ASTE only C$^{18}$O image cube
as discussed in the last subsection. In the lower velocity region (6.6 -- 7.4 km s$^{-1}$)
the envelope emission extends over the entire region.

\begin{figure}
\epsscale{1.0}
\plotone{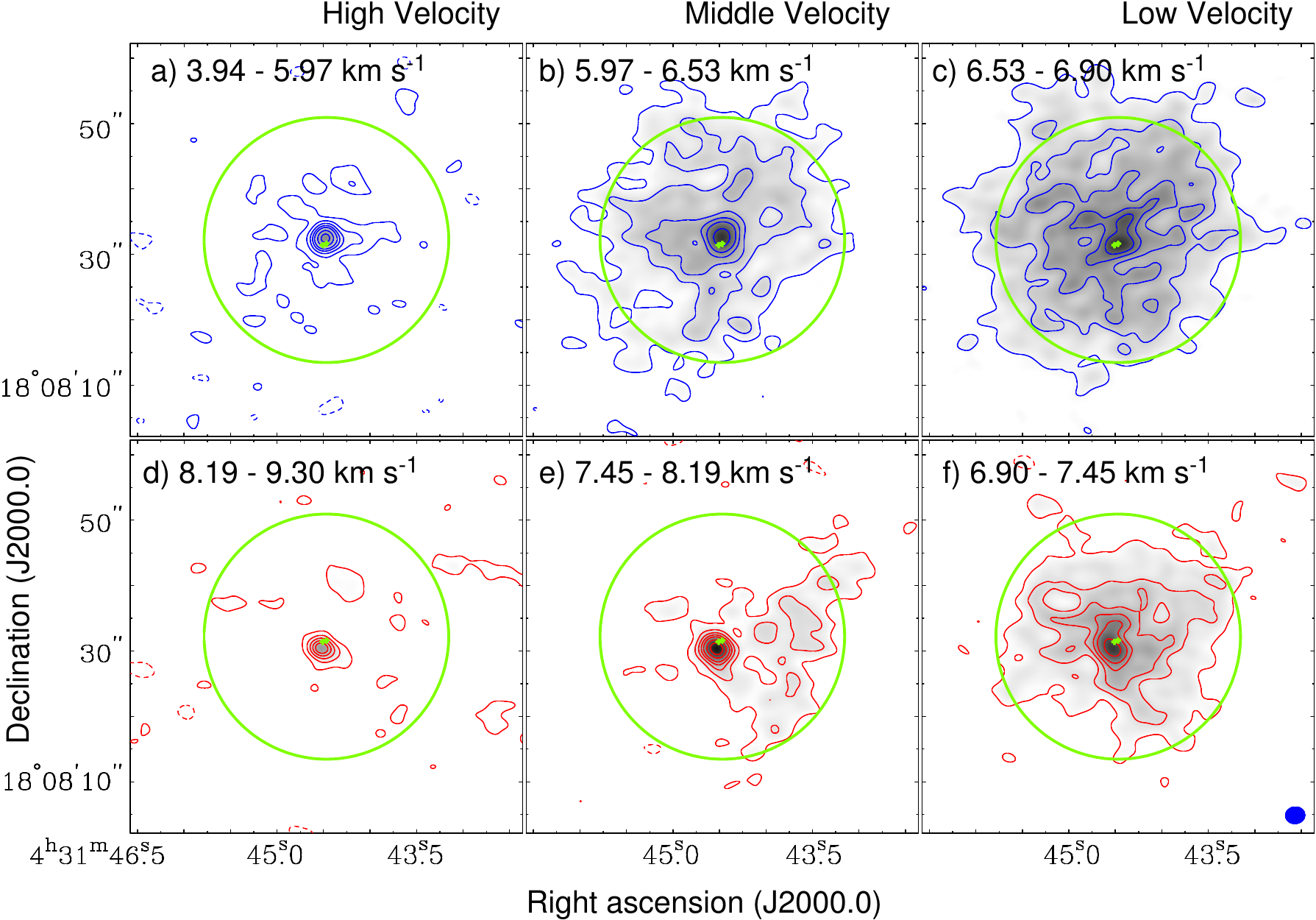}
\caption{SMA+ASTE maps of the high-, middle-, and low-velocity
blueshifted (blue contours) and redshifted (red contours) C$^{18}$O (3--2) emission
in L1551 NE. Integrated velocity ranges are labeled at the top-left corners of the panels.
Contour levels are in steps of 3$\sigma$ until 15$\sigma$, and then in steps of 5$\sigma$,
where 1$\sigma$ noise levels are 0.178 K, 0.339 K, 0.420 K, 0.242 K, 0.307 K, and
0.339 K in panels (a), (b), (c), (d), (e), and (f), respectively.
Symbols are the same as those in Figure \ref{mom0s}.
\label{comb3}}
\end{figure}

To highlight these velocity structures in the SMA+ASTE image cube,
we integrated the velocity channel maps over the high-velocity, middle-velocity,
and the low-velocity blueshifted and redshifted regions, and the resultant images
are shown in Figure \ref{comb3}. In the high-velocity region, compact blueshifted
and redshifted C$^{18}$O emission located to the north and south of the protostellar binary,
which originates from the Keplerian CBD, are seen. In the middle-velocity
region, as well as the central disk component, an extended envelope component, with the
blueshifted emission located predominantly to the east and the redshifted emission
to the west, are seen. As we discussed in the last subsection, one interpretation
of this velocity structure is the dispersing gas motion in the flattened envelope.
In the low-velocity region both the blueshifted and redshifted emission spread
over almost the entire field of view.
We note, however, that the blueshifted and redshifted
emission peaks in the central compact emission are located slightly
to the west and east of the protostellar binary, respectively. The sense of this velocity structure
is opposite to that of the extended envelope seen in the middle velocity range.
This velocity structure is also seen in the SMA only image cube and the centroid
velocity of this velocity feature matches with that of the CBD.
As discussed in our previous paper \cite{tak13},
one interpretation for this velocity structure is an infalling gas motion
in the inner envelope toward the central CBD.
Such coexistence of inner infall and outer dispersing motions has also been identified in
an envelope surrounding a T-Tauri star DG Tauri \cite{kit96}.

Figure \ref{combpv} (left) shows the P-V diagrams of the combined SMA+ASTE image cube
of the C$^{18}$O line along the major (upper panel) and minor axes (lower) of the CBD.
For comparison, Keplerian rotation curves derived from the SMA data \cite{tak12} are drawn
in black lines, and the centroid velocity of the CBD and the envelope in vertical dashed lines.
In the P-V along the major axis, the central compact component, CBD, exhibits a velocity gradient
consistent with the Keplerian rotation. On the other hand, the extended envelope component
does not show any clear velocity gradient along the major axis, and the offset of the centroid
velocity of the envelope from that of the CBD is clearly seen. These results indicate
that the Keplerian rotation of the central CBD is discontinuous to the outer envelope.
In the P-V diagram along the minor axis, the central component exhibits a wide velocity width
but no velocity gradient, as expected from the Keplerian rotation.
On the other hand, along the minor axis
the emission ridges in the envelope component are located
predominantly to the northeast in the blueshifted velocities, and to the southwest
to the redshifted velocities, as seen in the middle velocity ranges of Figure \ref{comb3}.
This velocity structure can be interpreted as the dispersing gas motion
in the flattened envelope. We will interpret these detected velocity structures
of the combined SMA+ASTE C$^{18}$O image cube with our toy model below.

\section{Analyses}
\subsection{Toy Model for the Combined SMA+ASTE Image Cube}

\begin{figure}
\epsscale{0.7}
\plotone{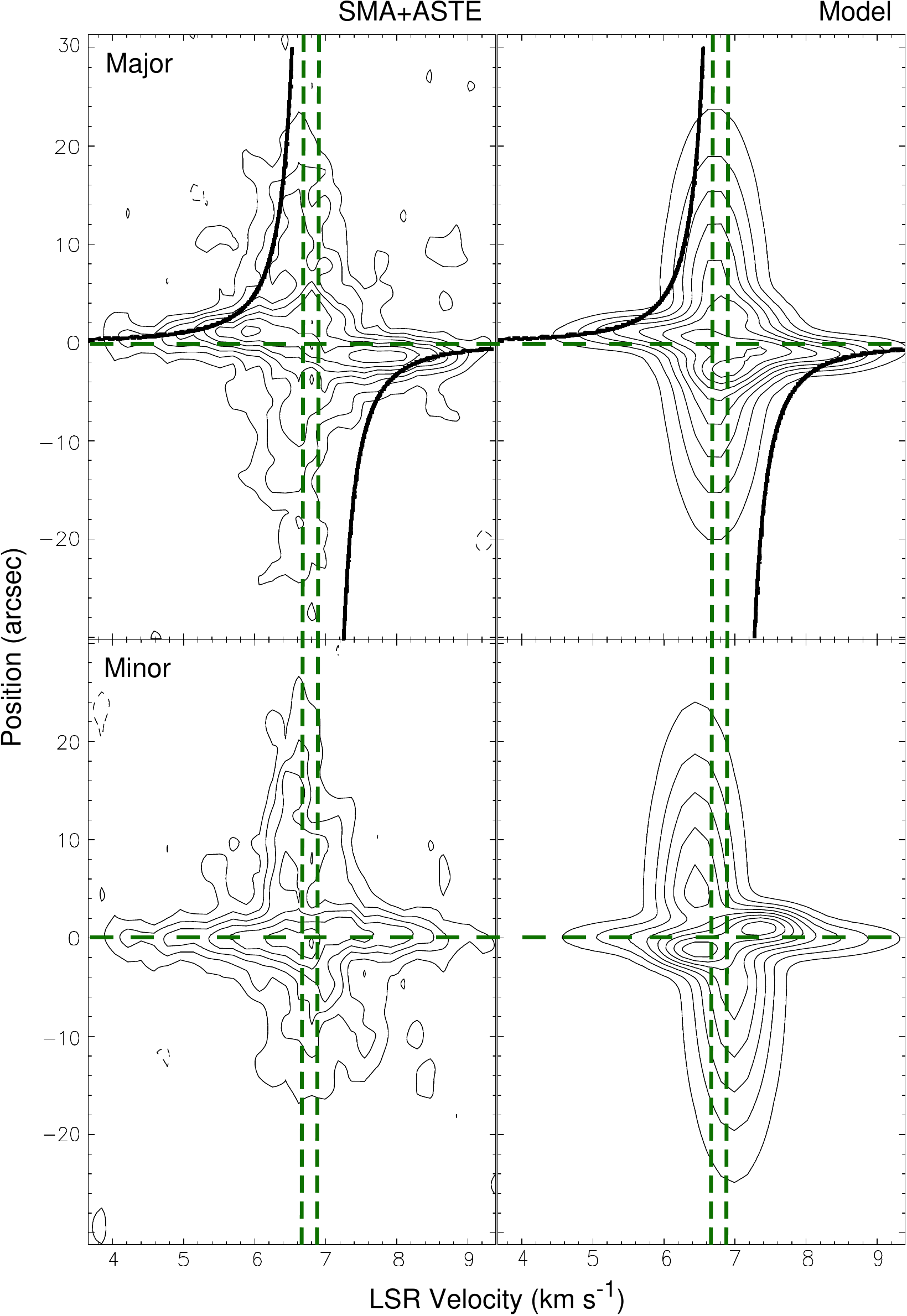}
\caption{Left: SMA+ASTE P-V diagrams of the C$^{18}$O (3--2) emission in L1551 NE
along the major (P.A.=167$\degr$; upper panel) and minor axes (P.A.=77$\degr$; lower)
of the Keplerian CBD. Contour levels are in steps of 2$\sigma$ (1$\sigma$ = 0.598 K).
Horizontal dashed lines denote the position of the protostellar binary (Source A), while
two vertical dashed lines the systemic velocities of the extended envelope component
($V_{LSR}$ = 6.7 km s$^{-1}$) and the disk ($V_{LSR}$ = 6.9 km s$^{-1}$).
Solid black curves show the Keplerian rotation curve of the disk
($i.e.$, $M_{\star}$ = 0.8 $M_{\odot}$ and $i$ = 62$\degr$).
Right: P-V diagrams of our toy model. Contour levels and symbols are
the same as those in the left panel.
\label{combpv}}
\end{figure}

The combined SMA+ASTE C$^{18}$O image cube reveals a sign of
an outer dispersing envelope motion, possible inner infalling motion,
and Keplerian rotation in the central CBD. To interpret these results
quantitatively, we constructed a toy model of the envelope and disk assuming that the disk
and envelope are geometrically thin and co-planar. Our model is essentially the same as that
by Takakuwa et al. (2013), which reproduces the velocity structures of the SMA only image cube,
but now the outer dispersing envelope component is added to reproduce the velocity
structure of the extended envelope component traced by ASTE.
As a model moment 0 map, we adopted a combination of two 2-dimensional Gaussians,
which are derived from the Gaussian fitting to the observed SMA+ASTE moment 0 map
(Figure \ref{mom0s} bottom-right).
The velocity structures in the inner component traced by the SMA are described as,
\begin{mathletters}
\begin{eqnarray}
v_{rot}(r)=\sqrt{\frac{GM_{\star}}{r}} \hspace{0.5cm} {\rm and} \hspace{0.5cm} v_{rad} (r) = 0 \hspace{0.5cm} {\rm for} \hspace{0.5cm} r \leq r_{kep}, \\
v_{rot}(r)=0 \hspace{0.5cm} {\rm and} \hspace{0.5cm}  v_{rad} (r) = v_{inf} \hspace{0.5cm} {\rm for} \hspace{0.5cm} r > r_{kep},
\end{eqnarray}
\end{mathletters}
\noindent where $r$ is the radius, $G$ the gravitational constant,
$M_{\star}$ the mass of the protostellar binary,
$v_{rot} (r)$ and $v_{rad} (r)$ are the rotational and radial velocities, respectively,
$r_{kep}$ the outermost radius of the Keplerian rotation, and $v_{inf}$ an adopted constant infall velocity
beyond $r_{kep}$.
As discussed by Takakuwa et al. (2013),
we adopt $M_{\star}$ = 0.8 $M_{\odot}$, $r_{kep}$ = 300 AU, and $v_{inf}$ = -0.6 km s$^{-1}$,
and the disk inclination angle $i = 62\degr$.
The velocity structure of the outer envelope traced by ASTE is expressed as,
\begin{equation}
v_{rot} (r)=0 \hspace{0.5cm} {\rm and} \hspace{0.5cm}  v_{rad} (r) = v_{disp} \hspace{0.5cm} {\rm for} \hspace{0.5cm} r > r_{disp},
\end{equation}
where $r_{disp}$ and $v_{disp}$ are the innermost radius of the dispersing envelope
and the dispersing velocity, respectively.
$r_{disp}$ = 700 AU is adopted, inferred from the comparison
between the SMA only and SMA+ASTE moment 0 maps.
$v_{disp}$ can be roughly estimated from the velocity gradient seen in the combined P-V diagram
along the minor axis as shown in Figure \ref{combpv}.
The velocity shift of the emission ridge of the envelope across the minor axis
is $\sim$0.5 km s$^{-1}$, centered on the centroid velocity of 6.7 km s$^{-1}$,
and at an inclination of $i$ $\sim$ 62$\degr$ the dispersing velocity along the flattened envelope is
$v_{disp}$ $\sim$(0.5 km s$^{-1}$ / 2) / sin 62$\degr$ $\sim$0.3 km s$^{-1}$.
Note that the centroid velocity of the envelope component is offset by $\sim$-0.2 km s$^{-1}$
from that of the CBD, and in our model this offset of the centroid velocity is
incorporated. The internal velocity dispersion $\sigma_{gas}$ = 0.4 km s$^{-1}$, which reproduces
the velocity channel maps of the central CBD \cite{tak12},
is adopted throughout the entire components.

\begin{figure}
\epsscale{0.95}
\plotone{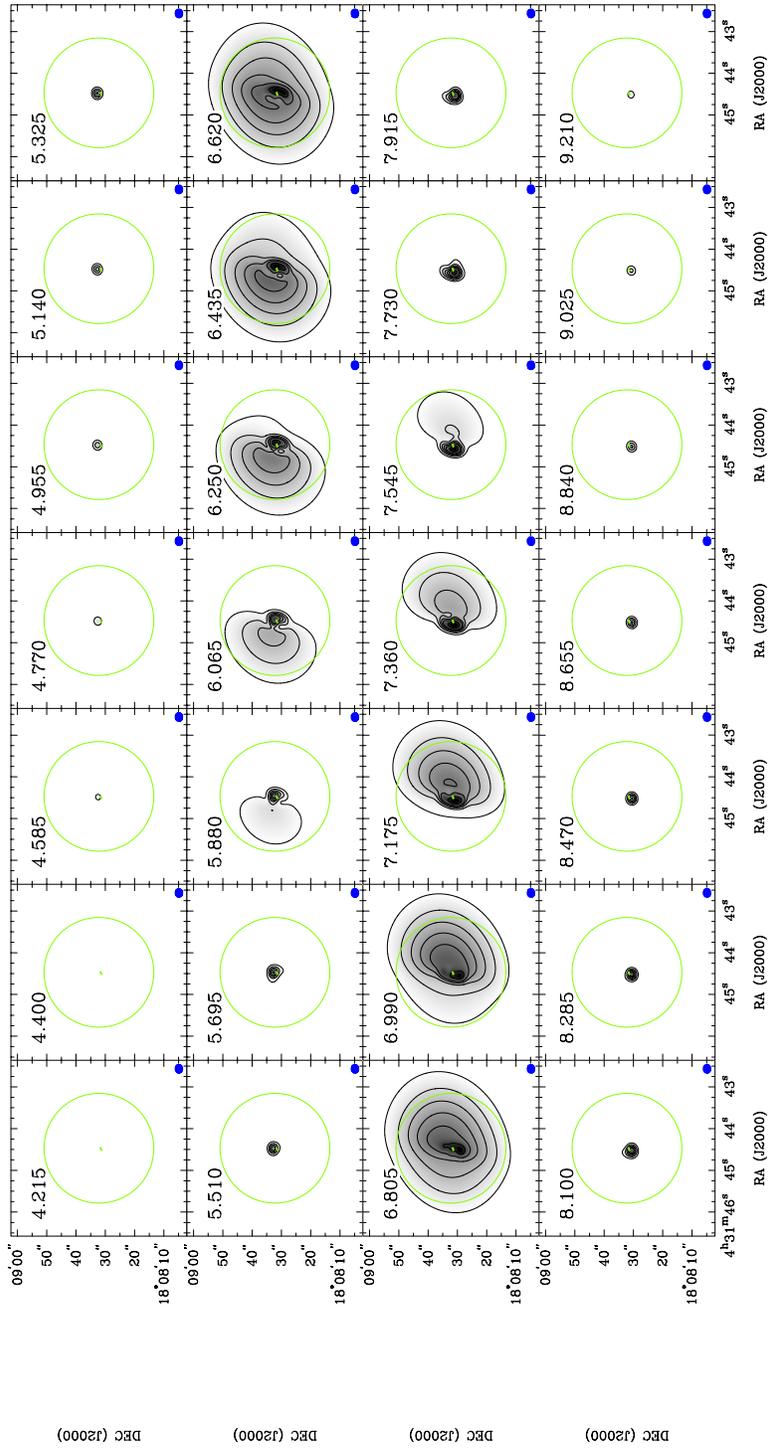}
\caption{Velocity channel maps of our toy model of the C$^{18}$O (3--2) line in L1551 NE
(see texts for details.).
Velocity ranges, contour levels, and symbols are the same as those in Figure \ref{combch}.
\label{modelch}}
\end{figure}

Figure \ref{modelch} shows the velocity channel maps of our toy model.
In the high-velocity blueshifted
(4.6 -- 5.7 km s$^{-1}$) and redshifted ranges (7.7 -- 9.2 km s$^{-1}$)
the features of the Keplerian CBD are seen.
In the blueshifted range of 5.9 -- 6.3 km s$^{-1}$ the envelope component is
located to the east of the protostellar binary while in the redshifted range of
7.2 -- 7.5 km s$^{-1}$ to the west, reflecting the dispersing motion in the envelope.
In the lower-velocity range (6.4 -- 7.0 km s$^{-1}$) the envelope component is
extended both in the eastern and western sides, but the east (red) - west (blue)
velocity gradient arising from the infalling motion in the central component
is evident.
These characteristics reproduces primary features of the observed SMA+ASTE C$^{18}$O velocity
channel maps in L1551 NE (Figure \ref{combch}, \ref{comb3}).
Figure \ref{combpv} (right) shows the P-V diagrams of our toy model
along the major and minor axes. The model P-Vs can reproduce
primary features of the observed P-Vs, such as
the high-velocity component arising from the Keplerian disk,
and the extended envelope component which shows a velocity gradient
along the minor axis reflecting the dispersing motion, but no rotation
along the major axis. These results show that
the observed SMA+ASTE C$^{18}$O image cube in L1551 NE can be understood with
three distinct velocity components; a central Keplerian CBD
at a radius of $\lesssim$300 AU, a possible infalling region at 300 AU $<$ $r$ $<$ 500 AU,
and an outer dispersing envelope with a dispersing velocity of $\sim$0.3 km s$^{-1}$
at $r$ $\gtrsim$500 AU.

\subsection{Physical Properties of the Gas Components Identified with ASTE}

Our ASTE mapping observations of the L1551 NE region have revealed three distinct
gas components; $i.e.$, ENV, RED1, and RED2.
ENV is the component of the protostellar envelope
surrounding the protostellar binary system L1551 NE,
and exhibits the dispersing gas motion as discussed above.
RED1 and 2 are most likely the redshifted outflow components
driven from L1551 IRS 5, which are interacting with ENV.

To discuss the interaction between ENV and RED1 and 2 quantitatively,
physical parameters of these components were estimated as follows.
First, the moment 0 map of each component was made in the relevant tracers and
the velocity ranges listed in Table \ref{comphys}. As shown above, ENV is most clearly identified
in the C$^{18}$O (3--2) emission, and RED2 in the $^{13}$CO (3--2) emission.
We adopt the CS (7--6) emission to deduce the physical condition of RED1,
because in the CS emission RED1 can be identified as a distinct gas component
most easily, and the submillimeter CS emission must trace higher-temperature gas
associated with outflows. We also note that it is not straightforward to separate RED2
in the 3-dimensional space from the other lower-velocity components. We here identified RED2 as the 
$^{13}$CO emission component in the highest redshifted velocity range of 9.47 -- 12.53 km s$^{-1}$,
since in this velocity range RED2 is a distinct, highest redshifted component seen primarily in
the $^{13}$CO emission. After making the moment 0 maps of these distinct components,
two-dimensional Gaussian fittings to the moment 0 maps
were performed to derive the central positions, beam-deconvolved sizes along the major and minor axes,
and the total flux. For ENV a single two-dimensional Gaussian fitting to the moment 0 map
did not provide a satisfactory fit. Thus, for ENV
two-component, two-dimensional Gaussian fitting was performed, and the position was defined as the peak position
of the central Gaussian component, while the size was defined as the size of the outer Gaussian component.
The line widths of these components ($\equiv \Delta v$)
were derived from the FWHM values of the relevant
spectra toward the central positions.
Projected flow velocities of RED1 and RED2, $v_{flow}$, were defined as
the mean velocity at the center of the components with respect to the velocity of ENV (= 6.7 km s$^{-1}$).

Masses of the components ($\equiv M_{LTE}$) were derived from the total fluxes,
on the assumption of the LTE condition, optically-thin emission, and the molecular abundances
of $X_{C^{18}O}$ = 1.7 $\times$ 10$^{-7}$ \cite{cra04},
$X_{^{13}CO} / X_{C^{18}O}$ = 7.7 \cite{wr94},
and $X_{CS}$ = 6.8 $\times$ 10$^{-10}$ \cite{jor04}.
Since the peak brightness temperature ($\equiv T_B$) of the C$^{18}$O emission toward the protobinary position
exceeds $>$3.4 K, the excitation temperature ($\equiv T_{ex}$) must exceed $\gtrsim$9.0 K
as $J_{\nu}(T_{ex}) - J_{\nu}(T_{bg})$ must be higher than $T_B$. On the other hand, to excite
the submillimeter CS (7--6) emission a gas temperature as high as $\sim$60 K is required
\cite{ta07a,tak11}. Thus, $T_{ex}$ = 10 - 60 K is assumed as a probable range of $T_{ex}$.
Virial masses of the components ($\equiv M_{vir}$) were derived as;
\begin{equation}
M_{vir} = \frac{5DC_{eff}^2}{2G},
\end{equation}
where
\begin{equation}
C_{eff}^2 = \frac{\Delta v^2}{8 \ln 2}.
\end{equation}
In the above expressions $D$ denotes the geometrical mean of the sizes
along the major and minor axes of the components, and
$C_{eff}$ the effective sound speed of gas.
The isotropic velocity dispersion inside the components is assumed.
For RED1 and 2, the sizes ($\sim$4000 AU) are as small as the beam size
($\sim$3200 AU), and thus the line widths derived from the central spectra represent
the average velocity dispersions. For ENV, there is no discernible spatial variation
of the C$^{18}$O line width as shown in the P-V diagrams (Figures \ref{collipv}, \ref{astepvs}),
except for that arising from the central disk.
Thus, the assumption of the isotropic velocity dispersion inside these components is probably valid.

Momenta ($\equiv p$), internal gas energies ($\equiv E_{int}$), and kinetic energies ($\equiv E_{kin}$)
of the components were derived as;
\begin{mathletters}
\begin{eqnarray}
p^{RED1,2}=M_{LTE}v_{flow},\\
p^{ENV}=M_{LTE}v_{disp},
\end{eqnarray}
\end{mathletters}
\begin{equation}
E_{int} = \frac{3}{2}M_{LTE}C_{eff}^2,
\end{equation}
\begin{mathletters}
\begin{eqnarray}
E_{kin}^{RED1,2}=\frac{1}{2}M_{LTE}v_{flow}^2, \\
E_{kin}^{ENV}=\frac{1}{2}M_{LTE}v_{disp}^2.
\end{eqnarray}
\end{mathletters}

Table \ref{comphys} summarizes these derived physical parameters.
The size ($\sim$20000 AU) and LTE mass ($\sim$1-3 $M_{\odot}$)
of ENV are typical of protostellar envelopes \cite{and00,mye00,tak03}.
The virial mass of ENV is comparable to or slightly larger than the
LTE mass plus the central protobinary mass. From the comparison
between the virial and LTE masses it is not straightforward to
ascertain whether ENV is gravitational bound or not. On the other hand,
the identification of the dispersing gas motion in ENV implies that
ENV is gravitationally unbound.
The outflow components of RED1 and 2 are characterized with their compact sizes
($\lesssim$4000 AU) and wide line widths ($\sim$1--2 km s$^{-1}$).
The LTE masses of the outflow components are likely lower
than the corresponding virial masses,
although for RED1 only the upper limit of the virial mass is obtained.
These results indicate that RED1 and 2 are not gravitationally-bound gas
condensations, consistent with our interpretation that RED1 and 2 are
outflowing gas components driven from L1551 IRS 5.
The summed flow momentum of RED1 and 2 is comparable to that
of the momentum of the dispersing motion in ENV.
Furthermore, the net internal+kinetic energy of these outflow components
is also comparable to the internal+kinetic gas energy of ENV.
These results imply that RED1 and 2
can exert significant dynamical impacts on ENV,
and can induce the observed dispersing gas motion of ENV through the interactions.

\section{Discussion}
\subsection{Dispersing Envelope around the Circumbinary Disk in L1551 NE}

\begin{figure}
\epsscale{0.90}
\plotone{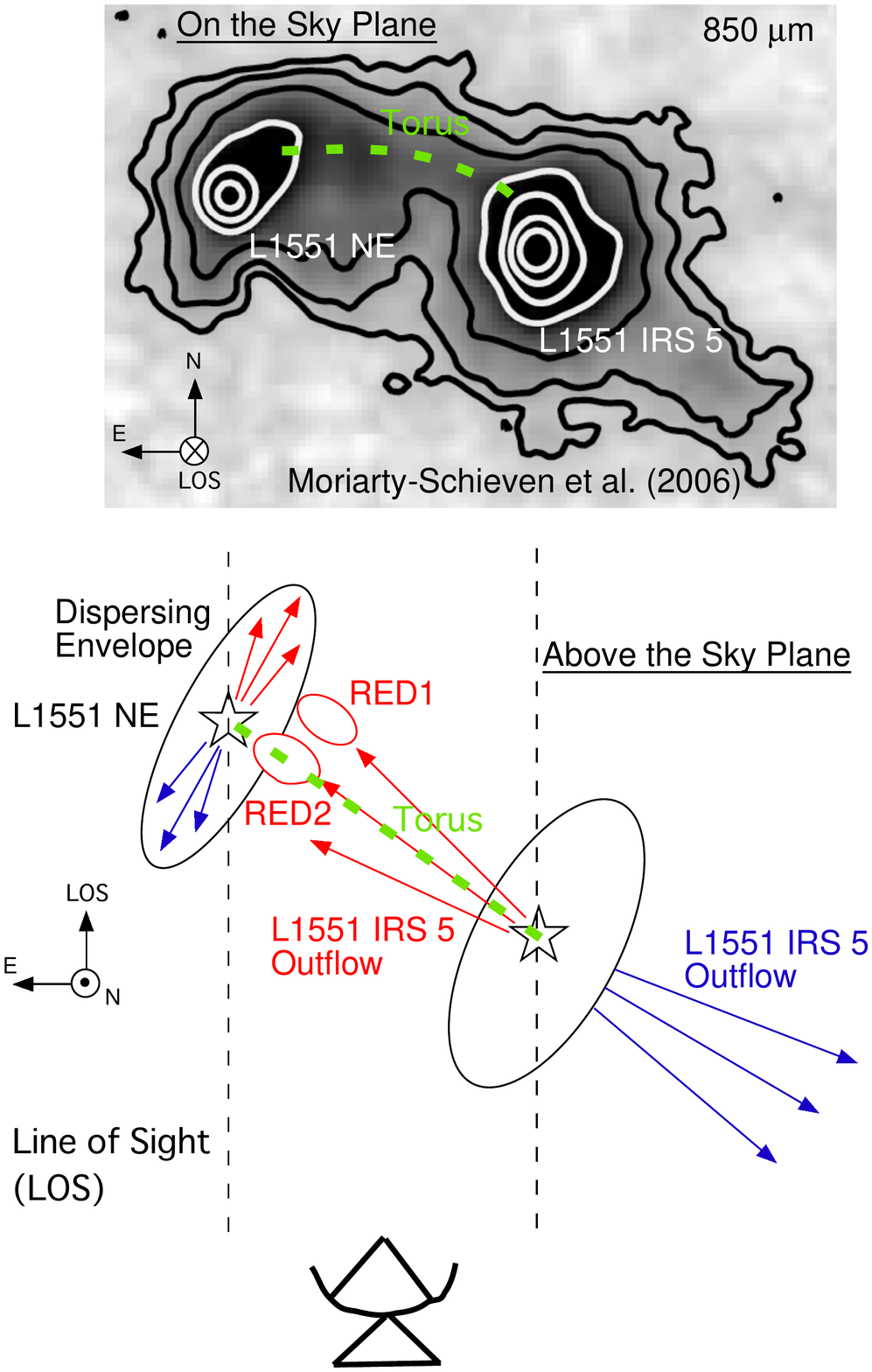}
\caption{Schematic illustration of the L1551 IRS 5 / NE system. The upper figure shows
the 850 $\micron$ dust-continuum image of the L1551 region taken with SCUBA \cite{mor06}.
The lower illustration shows a schematic,
anticipated view of the L1551 IRS 5 / NE system viewed from the north.
Green dashed curves denote the ``torus'' structure of the continuum emission connecting
between L1551 IRS 5 and NE.
\label{scheme}}
\end{figure}

Our new ASTE observations and combining the ASTE data with our previous SMA data in L1551 NE
have revealed that the protostellar envelope around the Keplerian CBD is being dispersed.
The dispersing envelope does not show any rotating gas motion in contrast with
the Keplerian rotation in the CBD,
and the centroid velocity of the envelope is $\sim$0.2 km s$^{-1}$
blueshifted with respect to that of CBD. These results indicate that
the protostellar envelope is kinematically distinct from the central CBD.
The outflow components from L1551 IRS 5, as seen most prominently in the
submillimeter CS (7--6) line (RED1) and the high-velocity redshifted
$^{13}$CO (3--2) line (RED2), appear to collide and interact with the protostellar envelope.
The net momentum, and kinetic and internal energy of these outflow components
are comparable to those of ENV, suggesting that
the outflow collisions can enforce significant dynamical impacts on the envelope.
A natural interpretation of these results is that the protostellar envelope in L1551 NE
is being disrupted through the interaction with the outflows driven from L1551 IRS 5.
Figure \ref{scheme} (bottom) shows a schematic, inferred configuration of the L1551 region
viewed from the north, in parallel with the SCUBA 850 $\micron$ dust-continuum image \cite{mor06}
as a representative projected image onto the sky plane (top). In the SCUBA image
a torus-like continuum feature connecting the envelopes around L1551 IRS 5 and NE is seen
as delineated with a dashed green curve. This feature is drawn schematically in the bottom
as the northernmost feature.
Since the envelope around L1551 NE is being collided with the redshifted outflow
components driven from L1551 IRS 5 (RED1 and 2), the location of L1551 NE must be backward
with respect to the location of L1551 IRS 5 along the line of sight. As already described
the western side of the flattened envelope around L1551 NE is far-side and the
eastern side near-side. Thus the blueshifted emission to the east and the redshifted emission
to the west are interpreted as the dispersing gas motion in the flattened envelope.

Our ASTE single-dish observations of the $^{13}$CO (3--2), C$^{18}$O (3--2), CS (7--6),
and SO (7$_8$--6$_7$) lines, as well as SMA and ALMA observations of the $^{13}$CO (3--2)
and C$^{18}$O (3--2) lines \cite{tak12,tak13,tak14}, do not find
any evidence for molecular outflows driven from L1551 NE itself.
On the other hand, previous NIR observations of L1551 NE found collimated [Fe II] jets
driven from Source A as well as a reflection nebula with its apex close to the position of Source B \cite{rei00,hay09}.
Molecular outflows driven from L1551 NE have also been reported
in the $^{12}$CO (3--2; 2--1; 1--0) lines \cite{mo95a,mor06,sto06},
although it is not straightforward to disentangle the outflows from L1551 IRS 5 and NE
because of the close overlaps of the two outflows.
Our ASTE observations show that interactions
with the powerful L1551 IRS 5 outflows have greater impact on the disruption of the envelope
than the outflow driven from L1551 NE itself. This is in contrast with previous
findings of interactions between protostellar envelopes and outflows or stellar winds driven from the sources
embedded in the envelopes \cite{kit96,mom96,fue02,tak03,arc06,tak06}.

Yokogawa et al. (2003) have suggested that the interactions with the outflows driven
from L1551 IRS 5 triggered the star formation in L1551 NE. We consider, however,
that the formation of the protostellar binary system L1551 NE has initiated before
the arrival of the IRS 5 outflows, because at present the protostellar envelope
is kinematically distinct and thus physically detached from the CBD and the
protostellar binary system.
The similar binary masses ($\sim$0.8 $M_{\odot}$ in L1551 NE and $\sim$0.5 $M_{\odot}$ in IRS 5),
envelope masses ($\sim$1 $M_{\odot}$), and the bolometric temperatures ($T_{bol}$ $\sim$90 K)
indicate that both L1551 IRS 5 and NE were born at a similar age. After their birth both L1551 NE
and IRS 5 grow simultaneously. The angular separation between L1551 IRS 5 and NE is $\sim$2$\farcm$5,
and assuming the transverse velocity of the outflows driven from L1551 IRS 5 is $\sim$10 km s$^{-1}$
\cite{mor06,sto06,wu09}, the transverse time for the IRS 5 outflow to reach the L1551 NE location
is $\sim$10$^{4}$ yr. This time scale is approximately consistent with
the estimated age of L1551 NE of 0.6--5.0$\times$10$^{4}$ yr \cite{mot01}.
After the outflows from L1551 IRS 5 have arrived at the protostellar
envelope around L1551 NE, the outflows start disrupting the envelope through
the interactions. The possible infalling motion in the inner envelope found with the SMA
may be a remnant gas motion before the initiation of the interactions with the IRS 5 outflows.
Such a remnant infalling gas motion embedded in the dispersing envelope has also been
identified in a T-Tauri star DG Tauri \cite{kit96}.

The envelope dispersion via the interactions with the IRS 5 outflows will impact the final fates of the
protostellar binary system of L1551 NE significantly,
as will be discussed in the next subsection.

\subsection{Implication of Envelope Dispersion for Binary Growth}

Binary stars with the primary masses comparable to that of the Sun
show a wide range of the binary mass ratios, and the distribution of the mass ratio ($\equiv q$)
is essentially flat from $q \sim$0.1 to 1 \cite{rag10,goo13}.
Physical mechanisms to reproduce such a wide range of binary mass ratios
have been controversial, and gas accretion in the CBDs
onto the protostellar binaries has been discussed as a mechanism to set the final binary mass ratios.
Smoothed particle hydrodynamic (SPH) simulations of circumbinary accretion show
that the majority of materials accrete onto the secondary, because the secondary orbits
further from the center of mass of the binary and thus sweeps more materials in the
CBD than the primary \cite{ba97a,ba97b,bat00}.
Our 3-D adaptive mesh refinement (AMR) hydrodynamic simulation \cite{mat07}
to reproduce the ALMA results of the CBD in L1551 NE also shows
that the secondary (Source B) accretes more than the primary (Source A) \cite{tak14}.
On the other hand, high-resolution
grid-based simulations by Ochi et al. (2005) and Hanawa et al. (2010) show that
primaries accrete more than secondaries.
Latest SPH simulations demonstrate that flows from the secondary to the primary
within the Roche lobes are a sensitive function of gas temperatures, and
in the case of the high temperature the primary accretes more than the secondary,
and vice versa \cite{you15}.

Our ASTE observations of an archetypal protostellar binary L1551 NE imply that
dispersion of protostellar envelopes, which replenish
CBDs with fresh materials, also needs to be taken into account
as a physical mechanism to set the binary masses and the ratios.
The present masses of Sources A and B in L1551 NE are estimated to be
$\sim$0.67 $M_{\odot}$ and $\sim$0.13 $M_{\odot}$,
respectively, and the mass of the CBD $\sim$0.026 $M_{\odot}$ \cite{tak12}.
The mass of the possible infalling component around the CBD is even smaller
($\sim$0.0023 $M_{\odot}$) \cite{tak13}. Thus, even if all the amount of the
material in the CBD plus the infalling component
is accreted onto Source B, the mass ratio does not change much (from 0.19 to 0.24).
Replenishment of the material to the CBD from
the surrounding envelope is required to significantly change the mass and the mass ratio
of the protostellar binary system. Our ASTE observations show that the envelope is being dispersed, however,
and that further replenishment of materials from the envelope to the CBD will not be expected.
Therefore, even though L1551 NE is a Class I protostellar binary associated with the CBD
and the protostellar envelope, the mass and the mass ratio have been already close to the final values.
Physical processes before the start of the envelope dispersion through the interaction with the outflow driven from L1551 IRS 5,
such as the fragmentation of the initial pseudo-disk and subsequent accretion from the CBD
and the envelope, must be responsible to set the mass and mass ratio of the binary system.

Our ASTE observations of L1551 NE indicate that dispersion of natal envelopes surrounding
the protostellar binary systems is one of the key physical phenomena to stop the growth of the binary
and to determine the final masses and the mass ratios of the binary. As described in the last subsection, interactions
between protostellar envelopes and outflows, which causes the envelope dissipation, are ubiquitous.
Systematic studies of envelope dispersion surrounding the protostellar binaries with various
masses and mass ratios will be intriguing to understand the physical mechanism to set the final binary mass
ratios and the origin of the wide range of the binary mass ratios.

\section{Summary}
We have conducted single-dish mapping observations of dense gas around
an archetypal protostellar binary L1551 NE in
the C$^{18}$O ($J$=3--2), $^{13}$CO ($J$=3--2), CS ($J$=7--6), and the
SO ($J_N$=7$_8$--6$_7$) lines with ASTE. We have also combined
the ASTE mapping data in the C$^{18}$O line with the interferometric data
taken with the SMA. The main results are summarized below:

\begin{itemize}
\item[1.] All the four molecular lines are detected toward the protostellar position,
with multiple and/or non-Gaussian spectral shapes.
In particular, the C$^{18}$O spectrum consists of a narrow
($\Delta v \sim$0.7 km s$^{-1}$) component with a central velocity of
$V_{\rm LSR} \sim$6.7 km s$^{-1}$ and a wide ($\Delta v \sim$2.2 km s$^{-1}$) component
with a central velocity of $V_{\rm LSR} \sim$6.9 km s$^{-1}$.
The wide component originates from the compact circumbinary disk (CBD),
while the narrow component from the extended protostellar envelope.


\item[2.] The C$^{18}$O map in L1551 NE primarily traces a $\sim$20000-AU scale protostellar envelope (ENV),
elongated along the northwest to southeast direction, approximately parallel to the major axis of the central CBD.
Velocity channel maps of the C$^{18}$O line show that the blueshifted ($V_{\rm LSR} \sim$6.2--6.7 km s$^{-1}$)
emission is located to the northeast while the redshifted emission ($V_{\rm LSR} \sim$6.7-- 8.0 km s$^{-1}$)
to the southwest; $i.e.$, a velocity gradient of ENV along
the minor axis, while there is no detectable velocity gradient along the major axis.
The observed velocity gradient in ENV can be interpreted as a dispersing gas motion.
Velocity channel maps of the high-velocity redshifted ($V_{\rm LSR} \gtrsim$8.0 km s$^{-1}$) $^{13}$CO emission
show two $\lesssim$4000-AU scale redshifted components to the $\sim$40$\arcsec$ southwest of the central protostar (RED1)
and $\sim$20$\arcsec$ west (RED2). The submillimeter CS (7--6) emission
selectively traces RED1, suggesting high-temperature in this high-velocity
component.
Both RED1 and RED2 exhibit similar velocity gradients such that
the velocities become closer to the systemic velocity of ENV as the positions become closer to that of ENV.
These high-velocity redshifted components most likely trace the outflow components
driven from the neighboring protostellar source, L1551 IRS 5, which are colliding with ENV surrounding L1551 NE.

\item[3.] The combined ASTE+SMA map in the C$^{18}$O emission shows
that the $r \sim$300 AU scale CBD and $r \sim$700 AU scale infalling component
found in our previous SMA observations are surrounded by ENV
with a slightly blueshifted velocity ($\sim$0.2 km s$^{-1}$).
ENV does not show any rotational motion but dispersing gas motion
with an outward velocity of $\sim$0.3 km s$^{-1}$
in contrast to the inner gas components. These results indicate
that ENV is kinematically distinct from the inner gas components.

\item[4.] The net momentum, kinetic and internal energies of the redshifted outflow components
driven from L1551 IRS 5 are comparable to those of ENV, suggesting
that the outflow collisions can enforce significant dynamical impacts on ENV.
Thus, the dispersing motion in ENV is likely caused by the interactions
with the outflows driven from L1551 IRS 5.

\item[5.] Since ENV is being dispersed and the inner gas components do not have sufficient
materials to alter the mass and mass ratio of the protostellar binary of L1551 NE,
the current mass ($\sim$0.8 $M_{\odot}$) and the mass ratio ($\sim$0.19)
are close to the final values, even though L1551 NE is a young Class I protostellar binary.
Our ASTE+SMA observations of L1551 NE suggest that dispersion of natal protostellar envelopes
which can replenish the CBDs with new materials is one of the important
physical mechanisms to set the final binary mass ratios.
The interactions with the outflows from L1551 IRS 5 are unlikely to trigger the binary formation
in L1551 NE as suggested by previous studies, but suppress the further growth of the protostellar
binary system.

\end{itemize}



\acknowledgments
We would like to thank T. Matsumoto, T. Hanawa, J. Lim, N. Ohashi, and P. T. P. Ho
for their fruitful discussions. We are grateful to S. Ohashi for his support
during our ASTE observations. S.T. acknowledges a grant from
the Ministry of Science and Technology (MOST) of Taiwan
(MOST 102-2119-M-001-012-MY3) in support of this work.

\clearpage



\clearpage


\clearpage

\clearpage

\clearpage

\clearpage


\clearpage

\clearpage

\clearpage

\clearpage

\clearpage

\clearpage

\clearpage

\clearpage

\clearpage
\begin{deluxetable}{lll}
\tablewidth{0pt}
\tablecaption{Resolutions and Noise Levels of the C$^{18}$O (3--2) Image Cubes}
\tablehead{Parameter &ASTE &ASTE+SMA}
\startdata
Beam                     &23$\arcsec$ & 2$\farcs$85$\times$2$\farcs$45 (P.A. = -85$\fdg$3)\\
Velocity Resolution & 0.114 km s$^{-1}$  & 0.185 km s$^{-1}$ \\
Noise Level & 0.18 K & 0.60 K \\ 
\enddata
\end{deluxetable}

\clearpage
\begin{deluxetable}{llllccccccccc}
\tabletypesize{\scriptsize}
\rotate
\tablecaption{Physical Properties of the Gas Components in L1551 NE Identified with ASTE\label{comphys}}
\tablewidth{0pt}
\tablehead{
\colhead{Component} & \colhead{Tracer} & \colhead{RA\tablenotemark{a}} & \colhead{DEC\tablenotemark{a}} & \colhead{$V_{LSR}$} &
\colhead{Size\tablenotemark{a}} & \colhead{$\Delta v$\tablenotemark{b}} & \colhead{$v_{flow}$\tablenotemark{c}} &
\colhead{$M_{LTE}$\tablenotemark{d}} & \colhead{$M_{vir}$\tablenotemark{d}} & \colhead{$p$\tablenotemark{d}} &
\colhead{$E_{int}$\tablenotemark{d}} & \colhead{$E_{kin}$\tablenotemark{d}} \\
    &        &    &      &(km s$^{-1}$)    & (AU)         &(km s$^{-1}$) &(km s$^{-1}$) &($M_{\odot}$) &($M_{\odot}$) &($M_{\odot}$ km s$^{-1}$) &($\times$10$^{42}$ erg) &($\times$10$^{42}$ erg)}
\startdata
ENV &C$^{18}$O (3--2) &04 31 44.4 &18 08 36.4 &5.31 - 8.15 &22100$\times$18200 &0.72 &\nodata &1.3 - 3.1 &5.3 &0.38 - 0.93 &3.5 - 8.7 &1.14 - 2.78 \\
RED1       &CS (7--6)             &04 31 42.0 &18 08 06.0 &8.04 - 9.13 &$<$4000                     &0.97 &1.9        &0.02 - 0.83 &$<$1.9 &0.039 - 1.574 &0.10 - 4.23 &0.73 - 29.74 \\
RED2\tablenotemark{e}       &$^{13}$CO (3--2) &04 31 42.8 &18 08 35.0 &9.47 - 12.53 &4300$\times$4000     &2.0 &3.5 &0.02 - 0.05   &8.4 &0.069 - 0.169    &0.42 - 1.04 &2.4 - 5.9 \\
\enddata

\tablenotetext{a}{Derived from two-dimensional Gaussian fittings to the moment 0 maps of the individual components. For
ENV, two-component, two-dimensional Gaussian fitting is performed, and the position is defined as the peak position
of the central Gaussian component, while the size is defined as the size of the outer Gaussian component.}
\tablenotetext{b}{FWHM line widths of the spectra toward the center of the components.}
\tablenotetext{c}{Mean velocity at the center of the components minus the velocity of the envelope component (= 6.7 km s$^{-1}$).}
\tablenotetext{d}{See texts for details.}
\tablenotetext{e}{It is not straightforward to unambiguously separate RED2 from the other lower-velocity components
in the $^{13}$CO emission. Here RED2 is defined from the isolated component in the velocity higher than 9.47 km s$^{-1}$.}
\end{deluxetable}

\end{document}